\documentclass{aa}
\usepackage[varg]{txfonts}
\usepackage{graphicx,aas_macros,natbib}
\usepackage{subfigure,aas_macros}

\def\stacksymbols #1#2#3#4{\def\theguybelow{#2}
        \def\verticalposition{\lower#3pt}
        \def\spacingwithinsymbol{\baselineskip0pt\lineskip#4pt}
        \mathrel{\mathpalette\intermediary#1}}
\def\intermediary #1#2{\verticalposition\vbox{\spacingwithinsymbol
        \everycr={}\tabskip0pt
        \halign{$\mathsurround0pt#1\hfil##\hfil$\crcr#2\crcr
                \theguybelow\crcr}}}
\def\lta{\stacksymbols{<}{\sim}{2.5}{.2}}
\def\gta{\stacksymbols{>}{\sim}{2.5}{.2}}

\newcommand{\pcc}{cm$^{-3}$}

\newcommand{\Pt}{$P_{\rm t}\ $}

\begin{document}

\title{Constraining turbulence and conduction in the hot ICM \\through density perturbations}

\author{M. Gaspari\thanks{E-mail: mgaspari@mpa-garching.mpg.de} \and E. Churazov} 
\institute{Max Planck Institute for Astrophysics, Karl-Schwarzschild-Strasse 1, 85741 Garching, Germany
}

\abstract{
Turbulence and conduction can dramatically affect the evolution of baryons in the universe; 
current constraints are however rare and uncertain. 
Using 3D high-resolution hydrodynamic simulations, tracking both electrons and ions, we study the effects of turbulence and conduction in the hot intracluster medium. We show how the power spectrum of the gas density perturbations ($\delta=\delta\rho/\rho$) can accurately constrain both processes. 
The characteristic amplitude of density perturbations is linearly related to the strength of turbulence,
i.e.~the 3D Mach number, 
as $A(k)_{\delta, \rm max} = c\, M$,
where $c\simeq0.25$ for injection scale of 500 kpc.
The slope of $A_\delta(k)$ in turn reflects the level of diffusion, dominated by conduction.
In a non-conductive medium, subsonic stirring motions advect density with a similar nearly Kolmogorov cascade,
$E_\delta(k)\propto k^{-5/3}$. Increasing conduction (parametrized via the magnetic suppression $f=10^{-3}\rightarrow1$) progressively steepens the spectrum towards the Burgers-like regime, $E_\delta(k)\propto k^{-2}$. The slope is only weakly dependent on $M$. 
The turbulent Prandtl number defines the dynamic similarity of the flow;
at scales where $P_{\rm t} \equiv t_{\rm cond}/t_{\rm turb} < 100$, the power spectrum  develops a significant decay, i.e.~conduction stifles turbulent regeneration. The transition is gentle for highly suppressed conduction, $f\le10^{-3}$, while sharp in the opposite regime. For strong conductivity ($f\ge0.1$), $P_{\rm t}\sim100$ occurs on spatial scales larger than the injection scale, globally damping density perturbations   
by a factor of 2$\,$-$\,$4, from large to small scales.
The velocity spectrum is instead not much affected by conduction.
The $f\ge0.1$ regime should also affect the appearance of X-ray images, in which Kelvin-Helmholtz and Rayleigh-Taylor 
rolls and filaments are washed out. 
In a stratified system, perturbations are characterized by a mixture of modes: weak/strong turbulence induces higher isobaric/adiabatic fluctuations,
while conduction forces both modes towards the intermediate isothermal regime. 
We provide a general analytic fit which is applied to new deep {\it Chandra} observations of Coma cluster. The observed spectrum is best consistent with strongly suppressed effective isotropic conduction, $f\simeq10^{-3}$, and mild subsonic turbulence, $M\simeq0.45$ (assuming injection scale at $\sim$250 kpc). The latter implies $E_{\rm turb}\simeq0.11\, E_{\rm th}$, in agreement with cosmological simulations and line-broadening observations.
The low conductivity 
corroborates the survival of sharp features in the ICM (cold fronts, filaments, bubbles), 
and indicates that cooling flows may not be balanced by conduction.}

\authorrunning{Gaspari \& Churazov} 
\titlerunning{Constraining turbulence and conduction in the hot ICM} 

\maketitle

\section{Introduction}\label{s:intro}

The intracluster medium plays central role in the evolution of baryons in the universe.
The ICM is the hot plasma filling the gravitational potential of galaxy clusters, the largest virialized
structures in the universe. Since most of the cluster baryons reside in the ICM ($\sim$85 percent),
this gaseous medium represents the crucible out of which essential astrophysical structures condense. 
It is often assumed that the ICM settles in hydrostatic equilibrium after the initial cosmological collapse in the potential
well of the cluster. However, the ICM plasma is a remarkably dynamic entity, continuously perturbed by mergers, feedback processes (AGN, supernovae), galaxy motions, and cosmological accretion, all shaping a chaotic and turbulent atmosphere.

Current X-ray observations have in general hard time detecting surface brightness (SB$_{\rm x}$) fluctuations,
due to the significant level of Poisson noise dominating on small scales (several tens kpc for nearby clusters) and projection
smearing effects. This has lead, throughout the past decades, to the common assumption that the ICM is a static entity, both in theoretical and observational work.
Only in recent time, few observational investigations have started to focus on the perturbations in the ICM, thanks to deep {\it Chandra} or {\it XMM} data. \citet{Schuecker:2004} found that at least $\sim$10 percent of the total ICM pressure in Coma hot cluster is in turbulent form. The spectrum of pressure fluctuations, in the range 40$\,$-$\,$90 kpc, appears to be  described by a Kolmogorov slope. Very recently, \citeauthor{Churazov:2012} (2012, 2013 -- in prep.) 
have analyzed very deep observations of Coma (650$\,\times\,$650 kpc; \S\ref{s:coma}). 
The characteristic amplitude of the relative density fluctuations reaches 5$\,$-$\,$10 percent, from small (30 kpc) to large scales (500 kpc), again resembling the Kolmogorov trend. 
\citet{Sanders:2012} also studied density/pressure perturbations in a cool-core cluster, AWM7, finding an amplitude of $\sim$4 percent, with a large-scale spectrum shallower than Kolmogorov.

Cosmological simulations (e.g.~\citealt{Norman:1999, Dolag:2005, Nagai:2007, Nagai:2013, Lau:2009, Vazza:2009, Vazza:2011, Borgani:2011}) indicate that subsonic chaotic motions are ubiquitous, with turbulent pressure support in the range 5$\,$-$\,$30 percent, from relaxed to merging clusters. On the other hand, large-scale simulations have severe difficulty in studying the details of perturbations, due to the limited resolution, the AMR derefinement, or SPH viscosity.
From a theoretical point of view, little attention has thus been paid to studying the role of density perturbations, and in particular the associated power spectrum, down to kpc scale. 
In idealized periodic boxes, \citet{Kim:2005} showed that isothermal turbulence produce a Kolmogorov spectrum,
progressively flattening with increasing Mach number. Even in the presence of weak magnetic fields, the power spectrum seems to retain the Kolmogorov slope, at least in ideal MHD simulations (\citealt{Kowal:2007}). 
In multiphase flows (e.g.~the interstellar medium), thermal instability generate a more complex nonlinear dynamics, inducing
high small-scale density perturbations, still increasing with $M$ (\citealt{Kissmann:2008,Gazol:2010}).

In the current work, we intend to carefully study the power spectrum (or better, the characteristic amplitude) of 3D density fluctuations, driven by turbulent motions (\S\ref{s:turb}) in a real galaxy cluster, Coma (\S\ref{s:coma}). 
This run will set the reference model.
We pay particular attention in modelling a realistic hot ICM plasma, as tracking both electrons and ions (2T; \S\ref{s:equip}), 
and avoiding restrictive assumptions on the equation of state (e.g.~isothermality).
Ideal hydrodynamics is however not enough to study a consistent evolution of an astrophysical plasma as the ICM. The very high temperatures ($\sim$$10^8$ K), combined with the low electron densities ($\sim$$10^{-3}$ cm$^{-3}$), warn that thermal conduction may have a profound impact in shaping density inhomogeneities (\S\ref{s:cond}).

The electron thermal conductivity of the ICM 
is a highly debated topic in astrophysics, and currently poorly (or not) constrained.
In the standard picture of a uniformly magnetized plasma, classic Spitzer conduction (\S\ref{s:cond}) is 
suppressed perpendicular to the $B$-field lines (by a factor $f\lta10^{-12}$), due to electron scattering
limited by the Larmor radius.
However, turbulent plasmas develop tangled magnetic fields with a chaotic topology.
According to \citet{Rechester:1978} and \citet{Chandran:1998}, after $\sim$30 times a random walk of the $B$-field coherence length $l_B$,
an electron in the ICM could be fully isotropized, leading to an effective isotropic suppression $f\sim10^{-2}-10^{-3}$,
still a substantially stifled conductive flux.
\citet{Narayan:2001} and \citet{Chandran:2004} further argued that, in a turbulent plasma, field lines can be chaotically tangled even on scales $< l_B$, 
possibly restoring the effective conductivity up to $f\sim0.1-0.4$ (the Spitzer value).

Past investigations have mainly focused on the role of conduction in balancing radiative losses.
In order to prevent the cooling catastrophe, the level of thermal conduction requires to be substantial, $f\gta0.1$,
or even impossible for several observed clusters, $f>1$ (\citealt{Kim:2003, Zakamska:2003, Voigt:2004}). 
Simulations also confirm the inefficiency of conduction (\citealt{Dolag:2004,Parrish:2009}), requiring other heating
mechanisms, as AGN feedback (\citealt{Churazov:2000,Churazov:2001,Ruszkowski:2002,Brighenti:2003, Gaspari:2011a,Gaspari:2011b,Gaspari:2012a,Gaspari:2013_rev}) or turbulent mixing (\citealt{Ruszkowski:2010,Ruszkowski:2011}).
Overall, observations lean towards highly suppressed conduction ($f\lta10^{-3}$), given the ubiquitous presence of cool cores, along with sharp temperature gradients linked to cold fronts (\citealt{Ettori:2000, Roediger:2013, ZuHone:2013}), X-ray cavities, or cold filaments (\citealt{Forman:2007, Sanders:2013_fil}). However, these constraints do not provide the effective isotropic conductivity in the {\it bulk} of the ICM; in fact, the magnetic field lines tend to naturally align perpendicular to the temperature gradient in a turbulent medium (\citealt{Komarov:2013}), hence strongly preventing the heat exchange between sharp fronts.  

Scope of this work is to provide, for the first time, a {\it global} constrain on the conductive and turbulent state of the ICM, instead of relying on local features. This will be possible exploiting the power spectrum of density perturbations.
After setting the physical and numerical framework (\S\ref{s:init}), we proceed step by step with controlled experiments (\S\ref{s:res}), 
assessing first the role of turbulence (weak, moderate, strong), and then,
gradually increasing the effective thermal conductivity.
The three features of the power spectrum unveil each a crucial aspect of the ICM state.
The normalization results to be linearly related to the turbulent Mach number. 
The slope of the spectrum steepens from Kolmogorov to Burgers trend, with rising conductivity. 
The decay/cutoff of the spectrum is provided by a key recurrent threshold, $t_{\rm cond}/t_{\rm turb}\lta100$ (\S\ref{s:res}). In \S\ref{s:disc}, we discuss important properties of the models,
and provide a simple model to assess the {\it effective} conductivity and turbulence.
We apply the prescription
to new very deep observations of Coma cluster, 
constraining the actual state of the hot ICM.
The results are summarized in \S\ref{s:conc}. In the Appendices, we compare different methods to calculate the spectrum, and analytically study the $\beta$-profile in Fourier space.
The increasing quality of future X-ray data will provide a big opportunity to exploit and perfect this new modelling, and hopefully to lead to high-precision measurements of the ICM (`ICMology').

\section[]{Physics \& Numerics} \label{s:init}
\subsection[]{Initial conditions: Coma galaxy cluster}\label{s:coma}
Hot galaxy clusters are optimal systems to study the effects of thermal conduction and turbulence, due to the fairly low ICM densities and the substantial level of dynamical activity.
The archetypal non-cool-core system is Coma cluster (Abell 1656). Given its proximity, brightness and flat X-ray core, it is ideal to study density perturbations. \citet{Churazov:2012} retrieved the characteristic amplitude of density fluctuations in Coma from deep {\it XMM} and {\it Chandra} observations, finding significant values up to 10 percent, while resolving scales of tens kpc.

\begin{figure} 
    \begin{center}
       \subfigure{\includegraphics[scale=0.55]{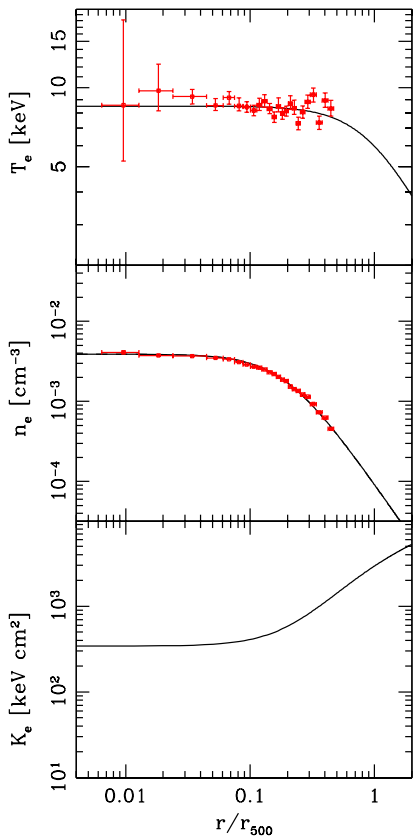}}
       \caption{Initial conditions for our reference hot galaxy cluster, Coma. From top to bottom panel: electron temperature, number density and entropy ($K=T_e/n_e^{2/3}$). The radius is normalized to $r_{500} \simeq 1.4$ Mpc. 
       We extracted the red data points from recent deep {\it XMM} observations, reaching $\sim0.5\ r_{500}$ (Lyskova et al.~2013, in prep.).
      \label{fig:Coma}}
     \end{center}
\end{figure}

In this study, we adopt Coma cluster as fiducial astrophysical laboratory, 
setting the density and temperature profile according to the most recent {\it XMM} observations (Fig.~\ref{fig:Coma}). An excellent fit to the radial electron density distribution is given by a single $\beta$-model profile:
\begin{equation}\label{nComa}
n_{\rm e} = n_{\rm e, 0} \left[1 + \left(\frac{r}{r_{\rm c}}\right)^2\right]^{-3 \beta/2},
\end{equation}
with central density $n_{\rm e, 0} = 3.9\times10^{-3}$ \pcc, core radius $r_{\rm c} = 272$ kpc, and $\beta = 0.75$.
In Appendix \ref{app:2}, we discuss the properties of the $\beta$-profile in Fourier space.
The gas temperature is roughly isothermal in the core, declining at large radii as observed for the majority of clusters (e.g.~\citealt{Vikhlinin:2006}):
\begin{equation}\label{TComa}
T = T_0\, \left[1 + \left(\frac{r}{r_{\rm t}}\right)^2\right]^{-0.45}, 
\end{equation}
where $T_0=8.5$ keV and $r_{\rm t}=1.3$ Mpc. The electron and ion temperature (\S\ref{s:equip}) are initially in equilibrium. The combined property of high temperature and low density sets a perfect environment to study the role of conduction, since the thermal diffusivity is $\propto T_{\rm e}^{5/2}/n_{\rm e}$ (\S\ref{s:cond}).
In addition, radiative cooling is ineffective, $t_{\rm cool}\sim2.5\,t_{\rm H}$.
The simulated system is initialized in hydrostatic equilibrium. The gas temperature and density allow to retrieve 
the gravitational acceleration (dominated by dark matter). The resulting potential is appropriate for a massive cluster in the $\Lambda$CDM universe, with virial mass $M_{\rm vir}\sim10^{15} M_\odot$ and $r_{\rm vir}\sim2.9$ Mpc ($r_{\rm 500}\sim1.4$ Mpc).

Since deep observations of density perturbations reach at most $0.5\, r_{\rm 500}$ (Fig.~\ref{fig:Coma}), we adopt a 3D box with a diagonal of $\sim$2.4 Mpc. As turbulence is volume filling and since we are interested in the power spectrum, the best numerical approach is
to use a fixed grid, without adaptive refinement. In fact, there is no trivial AMR criterium to apply due to the uniformly chaotic dynamics. Moreover, when the cube is de-refined by more than 50 percent, we found that that the density spectrum
has a significant decrease in power towards the small scales (producing a mock diffusivity), by over a factor of 2.
Based on these tests, we warn that using large-scale (cosmological) simulations will numerically steepen the slope of power spectra (density, velocity, etc.), as also shown by \citet{Vazza:2009}.
Albeit computationally challenging, we thus run all the models with fixed grid and high resolution of $512^3$ (considering the implemented physics).
We tested also $256^3$ runs, finding a very similar evolution and spectrum, though with double dissipation scale.
The simulations are thus in the convergence limit. 

The resolution is $\Delta x\,$$\sim\,$2.6 kpc, i.e.~roughly on the scale of the (unmagnetized) plasma mean free path. Going below this scale would formally require a kinetic approach. This also means that numerical viscosity is on the scale of the physical Spitzer viscosity, further corroborating the use of such resolution (\S\ref{s:visc}). The total evolution time is typically $\gta2$ eddy turnover times, roughly the statistical steady state after the turbulent cascade.
Boundary zones have Dirichlet condition, with value given by the large-scale radial profile. Inflow is prohibited, in order to avoid any spurious wave altering the dynamics inside the domain.

\subsection{Hydrodynamics} \label{s:hydro}

We use a modified version of the grid code FLASH4 (\citealt{Fryxell:2000}) in order to integrate the 3D equations of hydrodynamics for a 2 temperature (electron-ion; \S\ref{s:equip}) plasma, with the addition of turbulence (\S\ref{s:turb}) and electron thermal conduction (\S\ref{s:cond}): 
\begin{equation}\label{e:cont}
\frac{\partial\rho}{\partial t} + {\bf\nabla}\cdot\left(\rho {\bf v}\right) = 0
\end{equation}

\begin{equation}\label{e:mom}
\frac{\partial\rho {\bf v}}{\partial t} + \boldsymbol{\nabla}\cdot\left(\rho {\bf v} \otimes {\bf v}\right) + \boldsymbol{\nabla}{P} = 
\rho{\bf g}\, + \rho{\bf a}_{\rm stir}
\end{equation}

\begin{equation}\label{e:enei}
\frac{\partial\rho e_{\rm i}}{\partial t} + {\bf\nabla}\cdot\left(\rho e_{\rm i} \,{\bf v}\right) + P_{\rm i}  {\bf\nabla}\cdot{\bf v} =
\rho\mathcal{H}_{\rm i-e} 
\end{equation}

\begin{equation}\label{e:enee}
\frac{\partial\rho e_{\rm e}}{\partial t} + {\bf\nabla}\cdot\left(\rho e_{\rm e} \,{\bf v}\right) + P_{\rm e}  {\bf\nabla}\cdot{\bf v} =
\rho\mathcal{H}_{\rm e-i} - {\bf\nabla}\cdot \bf{F}_{\rm cond}
\end{equation} 

\begin{equation}\label{eos}
P_{\rm tot} = \left(\gamma -1\right)\,\rho (e_{\rm i} + e_{\rm e})
\end{equation}
where $\rho$ is the gas density, $\bf{v}$ the velocity, $e_{\rm i}$ and $e_{\rm e}$ the specific 
internal energy of ions and electrons, $P_{\rm tot}$ the total pressure (ions and electrons), 
$\gamma = 5/3$ the adiabatic index. The mean atomic weight of electrons and ions is $\mu_{\rm i}\simeq1.32$,
$\mu_{\rm e}\simeq1.16$, providing a total gas $\mu\simeq0.62$, appropriate for a totally
ionized plasma with $\sim$25\% He in mass. The atomic weight determines also the specific (isochoric) heat capacity,
$c_V=k_{\rm B}/[(\gamma-1)\,\mu m_{\rm p}]$.
For numerical consistency, the total energy ($e_{tot} = e_{\rm i} + e_{\rm e} + v^2/2$) equation, in conservative form, is also integrated.

In order to integrate the hyperbolic part of the hydrodynamics equations, we use a robust third order reconstruction scheme (PPM) in the framework of the unsplit flux formulation with hybrid Riemann solver (\citealt{Lee:2009}). Albeit computationally expensive, this setup keeps at minimum the numerical diffusivity. We tested different Riemann solvers (e.g.~HLLC, ROE), characteristic slope limiters (Min-Mod, Van Leer, Toro), and other parameters (e.g.~CFL number, interpolation order). They give comparable results, although we note that lower order reconstruction schemes (e.g.~MUSCL) are more diffusive and thus truncate the turbulent cascade on roughly two times larger scale.

\subsection[]{Turbulence driving}\label{s:turb}
Continuous injection of turbulence is modelled with a spectral forcing scheme that generates statistically stationary velocity fields (\citealt{Eswaran:1988, Fisher:2008, Gaspari:2013}). This scheme is based on an Ornstein-Uhlenbeck (OU) random process, analogous to Brownian motion in a viscous medium. The driven acceleration field is time-correlated, with zero mean and constant root mean square, an important feature for modelling realistic driving forces. In the OU process, the value of the gas acceleration at previous timestep $a^n$ decays by an exponential damping factor $f = \exp (-\,\Delta t / \tau_{\rm d})$, where $\tau_{\rm d}$ is the correlation time. Simultaneously, a new Gaussian-distributed acceleration with variance $\sigma_a^2=\epsilon^\ast/\tau_{\rm d}$ is added as
\begin{equation}\label{OU}
a^{n+1}_{\rm stir} = f \,a^n_{\rm stir} + \sigma_a\, \sqrt{1 - f^2} \; G^n,
\end{equation}
where $G^n$ is the Gaussian random variable, $\epsilon^\ast$ is the specific energy input rate, and $a^{n+1}_{\rm stir}$ is the updated acceleration. The six amplitudes of the acceleration (3 real and 3 imaginary) are evolved in Fourier space and then directly converted to physical space. In this approach, turbulence can be driven by stirring the gas on large scales and letting it cascade to smaller scales. This is an efficient approach as the alternative would involve executing FFTs for the entire range of scales, where the vast majority of modes would have small amplitudes. Since ICM turbulence is always subsonic, we impose a divergence-free condition on acceleration, through a Helmholtz decomposition in Fourier space.

The physical quantity of interest is the resultant 3D turbulent velocity dispersion, $\sigma_v$, which drives the ICM dynamics. The driving of turbulence is intentionally kept simple as our goal is not to consider any specific stirring source, but to keep the calculation fairly general. For example, the (combined) sources of turbulence may be major or minor mergers, galaxy motions, AGN feedback or supernovae. Observations (\citealt{Schuecker:2004, Churazov:2008, DePlaa:2012, Sanders:2013}) and simulations (\citealt{Norman:1999, Lau:2009, Vazza:2009, Vazza:2011, Gaspari:2012b, Borgani:2011} for a review) show that ICM turbulent energies are in the range few - 30 percent of the thermal energy, from very relaxed to merging clusters.

We test therefore three regimes of ICM turbulence, weak ($M\sim0.25$), mild ($M\sim0.5$), and strong ($M\sim0.75$), corresponding to a ratio of turbulent to thermal energy of 3.5, 14 and 31 percent ($E_{\rm turb} \simeq 0.56\, M^2 E_{\rm th}$). This is achieved by adjusting the energy per mode\footnote{Via simple dimensional analysis (\citealt{Ruszkowski:2010}), $\sigma_v\propto(N\, L\,\epsilon^\ast)^{1/3}$; the number of modes is typically $N<1000$.} $\epsilon^{\ast}$ and correlation time $\tau_{\rm d}$ ($\epsilon^{\ast}\sim5\times10^{-5}-10^{-3}$ cm$^2$ s$^{-3}$ and 200 Myr, respectively). As long as different choices of these parameters result in the same velocity dispersion, the dynamics of the flow remains unaffected. We stir the gas only on large scales, with typical injection peak $L\sim600$ kpc (in the last set of runs $L'\sim300$ kpc), letting turbulence to naturally cascade. 
This allows us to exploit the entire dynamic range of our box (512$^3$) and to better appreciate the effect of conduction, without being strongly affected by numerical diffusion.
Notice that in few $t_{\rm eddy}$, large-scale turbulence is not able to eject a substantial amount of mass outside the box. 
Since turbulence is kept subsonic, dissipational heating, which is proportional to $\sigma_v^3/L$, is also secondary. Turbulent diffusion can instead effectively flatten the global entropy gradient, especially in the non-conductive runs. We are nevertheless interested in the relative variations of $\delta\rho/\rho$, removing the underlying profile.

The characteristic time of turbulence is defined by the eddy turnover time at a given physical scale $l$ (e.g.~Fig.~\ref{fig:time1}, magenta line).
Extrapolating from the injection scale via the Kolmogorov scaling, $\sigma_v=\sigma_{v,{\rm inj}}\, (l/L)^{1/3}$, yields
\begin{equation}\label{e:tturb}
t_{\rm turb}= \frac{l}{\sigma_v}\sim\frac{L^{1/3}}{\sigma_{v,{\rm inj}}}\, l^{2/3}.
\end{equation}
Finally, we note turbulence can be expressed as a diffusion process acting on entropy
on sufficiently large scales, $\gta l$, 
albeit the equations are intrinsically hyperbolic. We define the effective turbulent diffusivity as
\begin{equation}\label{e:Dturb}
D_{\rm turb}= c_{\rm t}\, \sigma_v \,l.
\end{equation}
The transport of heat due to turbulent diffusion can be written as 
\begin{equation}\label{e:turb_heat}
\boldsymbol{\nabla} \cdot  {\bf F}_{\rm mix}  = -\boldsymbol{\nabla} \cdot  (D_{\rm turb} \rho T\boldsymbol{\nabla} s),
\end{equation}
where $s=c_V \ln(P/\rho^\gamma)$ is the entropy. We remark that turbulence diffuses entropy, while seeding perturbations
in density and temperature. In our analysis and discussion, we assume a diffusion constant $c_{\rm t} = 1$, but in real plasmas
this uncertain value could be much lower (\citealt{Dennis:2005} and references therein). 

\subsection{Thermal conduction} \label{s:cond}
In ionized plasmas such as the ICM, electrons conduct internal energy with a heating rate per unit volume given by
\begin{equation}\label{e:Hcond}
\boldsymbol{\nabla} \cdot  {\bf F}_{\rm cond} = -\,\boldsymbol{\nabla} \cdot  (\kappa\,  \boldsymbol{\nabla} T_{\rm e}),
\end{equation}
The thermal conductivity can be written as (\citealt{Spitzer:1962, Cowie:1977})
\begin{equation}\label{e:kcond1}
k \simeq f \,\,\frac{1.84\times 10^{-5}\, T_{\rm e}^{5/2}}{\ln \Lambda_{\rm ei}} \ \ \ {\rm [erg\, s^{-1}\, K^{-1}\, cm^{-1}]},
\end{equation}
where $\ln \Lambda_{\rm ei} = 37.8 +\ln [(T_{\rm e}/10^8\ {\rm K})\,(n_{\rm e}/10^{-3}\ {\rm cm^{-3}})^{-1/2}]$ is the Coulomb logarithm (the ratio of the largest to smallest impact parameter; e.g.~\citealt{Voigt:2004}), and $f$ is the magnetic suppression factor (\S\ref{s:MHD}).
The previous conductivity derives from the more significant expression
\begin{equation}\label{e:kcond2}
k \simeq (0.76\, f \,n_{\rm e}\, k_{\rm B})\, \lambda_{\rm e}\, v_{\rm e},
\end{equation}
which points out that the characteristic length scale and speed of conduction is the electron
mean free path $\lambda_{\rm e} \approx 10^4 T^2_{\rm e}/n_{\rm e}$ and the electron thermal speed $v_{\rm e}=(3 k_{\rm B} T_{\rm e}/m_{\rm e})^{1/2}$, respectively. Another important quantity is the isochoric diffusivity (cm$^2$ s$^{-1}$), defined as $D_{\rm Sp} = \kappa/c_{V, {\rm e}}\, \rho = \kappa/1.5\, n_{\rm e}\, k_{\rm B}\simeq 0.5\, v_{\rm e} \lambda_{\rm e}$. 

\subsubsection{Magnetic field suppression} \label{s:MHD}

The intracluster plasma is likely magnetized. Although the ICM magnetic field (few $\mu$G) appears dynamically unimportant compared to the thermal pressure, electrons and ions are anchored to the $B$-field lines.
The gyroradius or Larmor radius is many orders of magnitude lower than the Coulomb mean free path,
hence the charged particles can diffuse only along the $B$-field lines. 
On the other hand, a real atmosphere is always characterized by some degree of chaotic motions (e.g.~AGN/SN feedback, sloshing, mergers, galaxy motions). 
\citet{Ruszkowski:2010} showed through 3D MHD simulations that very weak subsonic turbulence always induces a tangled magnetic field with small coherence length ($l_B \lta 10$ kpc; figure 2). Faraday rotation measures also indicate small $l_B$ (e.g.~\citealt{Kim:1990}).
Therefore, on scales larger than the $B$-field coherence length, the average suppression due to anisotropic conduction can be parametrized with the so-called $f$ factor, as in our study.
If $l_B$ is instead very large (as in the idealized case of a parallel topology), this prescription is less accurate: parallel and perpendicular conductivity should be studied separately. The spectral analysis should be nevertheless not strongly affected by localized features (as fronts and filaments), weighing more volume-filling properties; Fourier spectra 
provide thus the conductivity in the {\it bulk} of the ICM.  

High-resolution MHD\footnote{In reality, only kinetic 3D simulations, solving the Vlasov equations, can exactly determine the effective $f$, tightly linked to magnetic instabilities, as firhose, mirror, etc. (\citealt{Schekochihin:2007}), which is out of reach for the current computing power.} simulations are required to assess the role of anisotropic conduction and will be tackled in a future work. 
Assuming a sufficiently tangled field, 
$f$ is geometrically expected to be $\langle cos^2 \theta \rangle \approx1/3$ ($\theta$ is the angle between the $B$ field and the $T$ gradient), as confirmed by turbulent MHD runs (\citealt{Ruszkowski:2010}). Microscopic effects and plasma instabilities can severely suppress the conductive flux down to $f\sim10^{-3}$ (\citealt{Rechester:1978, Chandran:1998}), although \citet{Narayan:2001} argue that in a highly tangled and chaotic field conduction may be restored to $f\sim0.2$ (\S\ref{s:intro}).
Considering these uncertainties, we test a wide range of $f$ values, 
ranging from the strongly to weakly suppressed regime, $f\sim10^{-3} - 1$.

From an observational point of view, deep data (e.g.~\citealt{Markevitch:2007}) show sharp contact discontinuities in the ICM, leading to the conclusion that conduction is severely suppressed, $f\sim10^{-3}$ (\citealt{Ettori:2000}). However, these estimates focus only on local features. Simulations of sloshing motions (e.g.~\citealt{ZuHone:2013}) show that the magnetic field tends to remain perpendicular to the temperature gradient (a natural outcome of the frozen-in property; \citealt{Komarov:2013}), with a strong suppression across the front. It is still unknown what is the effective {\it global} conductivity of the turbulent ICM (and hence global heat diffusion), which we aim to constrain with our spectral method, comparing simulated and observed density perturbations (\S\ref{s:comp}).

\subsubsection{Saturation}

Since the conductive flux depends also on the temperature gradient, electrons may conduct heat much faster than their thermal velocity in an unphysical way. This happens whenever the temperature scale height is smaller than the electron
mean free path, $l_T \equiv T/|\nabla T| \lta \lambda_e$. In this regime, the conductive flux saturates at a value given by (\citealt{Cowie:1977, Balbus:1982})
\begin{equation}\label{e:sat}
{\bf F}_{\rm sat} = - \,\alpha\, n_{\rm e}k_{\rm B}T_{\rm e}\, v_{\rm e}\, {\rm sgn}(\boldsymbol{\nabla}T),
\end{equation}
where $\alpha$ is an uncertainty factor representing microscopic processes in the magnetized plasma (as instabilities). Following the indication of \citet{Balbus:1986} based on plasma experiments, we set $\alpha\sim0.1$,
although the exact value has no great impact on the dynamics.
Saturation changes the nature of the equations, from parabolic to hyperbolic, yet we can define an effective diffusivity of the form $D_{\rm sat} = |\boldsymbol{\nabla}T/F_{\rm sat}|$.
Numerically, saturation is implemented via a smooth flux limiter on the diffusion coefficient: $D_{\rm cond} = (D_{\rm Sp}^{-2} + D_{\rm sat}^{-2})^{-1/2}$. We tested other types of limiters, as harmonic or min/max, without finding relevant differences. Saturation is only relevant in the unsuppressed regime, with typically less than 10 percent of zones saturated, while negligible for $f\lta0.1$ models. 

The final diffusivity is important to determine the characteristic timescale of conduction (see Fig.~\ref{fig:time1}, black line):
\begin{equation}\label{e:tcond}
t_{\rm cond} = \frac{l^2}{D_{\rm cond}}.
\end{equation}

In order to integrate the diffusion equation, we initially used an explicit flux-based scheme. However, the computational time becomes prohibitive since it is strongly limited by Eq. \ref{e:tcond}, allowing to integrate only few 100 Myr. It is thus essential to adopt the (unsplit) implicit solver, allowing for a fast yet accurate execution. The solver efficiently uses the HYPRE linear algebra package to solve the diffusion equation linked to electron thermal conduction
(cf. FLASH4 UG for the validation tests).
The associated boundaries are set in outflow or zero-gradient mode.

\subsection{2T plasma: electron - ion equilibration} \label{s:equip}
In astrophysical simulations, it is widely assumed that the plasma has one single temperature, $T_{\rm e}\approx T_{\rm i}$. However, this approximation is only good for a relatively cold medium. For hot clusters, especially non-cool-core systems, the ion-electron equilibration time due to Coulomb collisions can be $t_{\rm ei}\gta 50$ Myr. Since conduction operates on the Myr scale and {\it only} on electrons, it is important to
follow the evolution of {\it both} the electron and ion temperature (or internal energy; Eq. \ref{e:enei}-\ref{e:enee}).
The heat exchange rate (erg s$^{-1}$) between ions and electrons is given by
\begin{equation}\label{e:exc}
\mathcal{H_{\rm i-e}} =  \frac{c_{V,\rm e}}{t_{\rm ei}} (T_{\rm e} - T_{\rm i}), \\
\mathcal{H_{\rm e-i}} =  \frac{c_{V,\rm e}}{t_{\rm ei}} (T_{\rm i} - T_{\rm e}),
\end{equation}
where we choose the widely used Spitzer electron-ion equilibration time for a fully ionized plasma 
(\citealt{Huba:2009}): 
\begin{equation}\label{e:texc}
t_{\rm ei} = \frac{3\,k_{\rm B}^{3/2}}{8\sqrt{2\pi}\,e^4}\,\frac{(m_{\rm i} T_{\rm e} + m_{\rm e}T_{\rm i})^{3/2}}{(m_{\rm e} m_{\rm i})^{1/2}\,n_{\rm i}\, \ln \Lambda_{\rm ei}}.
\end{equation}
The equilibration time is dominated by $\propto (m_{\rm i}T_{\rm e})^{3/2}$. Therefore, in systems with characteristic
temperature $\lta6$ keV (and dense cores), the equilibration time is comparable or less than the unsuppressed conduction timescale ($\propto T^{-5/2}$), even on few kpc, $t_{\rm ei}\lta1$ Myr. Neglecting equilibration in the strongly conductive runs, the ions would be forced to be quickly isothermal, inducing spurious features. In the more realistic evolution, turbulent motions displace the ions before having time to fully equilibrate with electrons, leading to a gentler equilibration.
To our knowledge, this is the first attempt to study ICM turbulence and conduction with a 2T simulation, in analogy with high-energy physics studies.

\subsection{Viscosity} \label{s:visc}
Viscosity may in principle not be neglected when conduction operates, since both transport processes are intimately connected on the microscopic scale, and both altered by $B$-field instabilities and topology.
In the present work, we do not implement a direct physical viscosity, yet we point out two important arguments.
Since viscosity is the transport of momentum due to {\it ions}, while the conductive flux is associated with electrons,
viscous stresses are slower by at least 1$\,$-$\,$2 orders of magnitude compared with conduction (thermal $v_{\rm i} \simeq v_{\rm e}/43$), and should have a secondary role in damping density perturbations (see also \S\ref{s:res}). 
Further, we choose numerical resolution to be on the scale of the ion mean free path ($\lambda_{\rm i}\simeq\lambda_{\rm e}$);
the flow velocity is also comparable to the characteristic velocity of viscosity. 
This implies that numerical diffusivity approximately reflects Spitzer viscous diffusion ($\propto \lambda_{\rm i}\, v_{\rm i}$) -- the unsuppressed dynamic viscosity in Coma is $\eta_{\rm Sp} \simeq7.1\times10^{3}\,T_{8.5}^{5/2}$ g cm$^{-1}$ s$^{-1}$  
(cf.~\citealt{Reynolds:2005}). 
For example, the non-conductive spectrum in Fig.~\ref{fig:Ak1} indicates that the typical Reynolds number at injection is Re$_L \simeq (L/\lambda_{\rm T})^2\sim 500$, where the Taylor scale $\lambda_T$ marks the incipit of dissipation.
Comparing the simulated Taylor scale with the Spitzer value, $\lambda_T \simeq (\nu_{\rm Sp}\, L/\sigma_v)^{1/2}\sim 80$ kpc ($\nu_{\rm Sp} = \eta_{\rm Sp}/\rho$), we see that the grid acts as an effective viscosity with $f_\eta\sim1/4$.
In future, we will test the effects of
a varying and anisotropic physical viscosity.

\subsection{Power spectrum of density perturbations} \label{s:filter}
To study density perturbations,
we adopt the characteristic amplitude, instead of the power spectrum $P(k)$ (or energy spectrum $E(k)$), 
defined as
\begin{equation}
A(k) \equiv \sqrt{P(k)\, 4\pi k^3} = \sqrt{E(k)\, k},
\end{equation}
where $k= \sqrt{ k_x^2+k_y^2+k_z^2}$ (we typically use $l = 1/k$ in kpc units).
The characteristic amplitude is insightful, since the units are the same of the variable in real space. 
Since we are interested in the relative perturbations of density, $\delta \equiv\delta\rho/\rho$, $A(k)_\delta$ represents the typical level of fluctuations at a given scale $k$. The peak of $A(k)_\delta$ provides a good and simple estimate for the total amount of perturbations; the exact total variance can be computed integrating $P(k)\,4\pi k^2 dk$ over the whole range of scales (the difference may reach a factor of 1.4, depending on the power slope). 

We retrieve the relative density perturbations $\delta$, dividing $\rho$ by the background profile, $\delta = \rho/\rho_{\rm b} -1$. 
For each snapshot, we execute a best-fitting routine to compute the new underlying $\beta$-profile (e.g.~strong turbulent diffusion can lower the central density by a factor of 2 over few Gyr), minimizing the deviation between the data and the model.
We note that the $\beta$-profile removal affects only the spectrum on very large scales. In principle, $A(k)_\delta$ could be studied even without removing the background, since perturbations start to dominate on large $k$ (see Appendix \ref{app:2}).
The power spectrum of $\delta$ is finally computed with the `Mexican Hat' filtering (\citealt{Arevalo:2012}) instead of performing Fourier transforms, which can lead to spurious features due to the non-periodicity of the box (see Appendix \ref{app:1}).

\section[]{Results} \label{s:res}

\begin{table*}
\caption{Key properties of $A_\delta(k)$ for the simulated models: normalization (maximum), slope, and scale of significant decay.}
\label{table:1} 
\centering
\begin{tabular}{l | c c c c}
\hline\hline
{ } & Mach $\sim$ 0.25 & Mach $\sim$ 0.5 & Mach $\sim$ 0.75 & Mach $\sim$ 0.25, $L/2$  \\
\hline 
$A_\delta(k)$ normalization & & & & \\
$f = 0$ (hydro)            & 6.4\% & 11.7\% & 18.8\% & 5.3\% \\
$f = 10^{-3}$    & 6.4\% & 11.7\% & 18.8\% & 5.1\%  \\
$f = 10^{-2}$    & 6.0\% & 11.3\% & 18.6\% & 4.8\%  \\
$f = 10^{-1}$    & 4.6\% & 10.0 \% & 16.8\% & 3.5\% \\
$f = 1$             & 3.2\% & 9.1 \% & 15.4\% & 3.3\% \\
\hline
$A_\delta(k)$ slope\tablefootmark{\it{a}}  &   & & & \\
$f = 0$ (hydro)             & -1/3 & -1/5 & -1/5 & -1/5\\
$f = 10^{-3}$    & -1/3 $\rightarrow$\tablefootmark{{\it b}} -1/2 & -1/5 $\rightarrow$\tablefootmark{{\it b}} -1/2  & -1/5 $\rightarrow$\tablefootmark{{\it b}} -1/2 & -1/2\\
$f = 10^{-2}$    & -1/2 $\rightarrow$\tablefootmark{{\it b}} -4/5 & -1/3 $\rightarrow$\tablefootmark{{\it b}} -1/2  & -1/3 $\rightarrow$\tablefootmark{{\it b}} -1/2 & -1/2 \\
$f = 10^{-1}$    & -2/3 & -4/9 & -4/9 & -4/9  \\
$f = 1$             & -1/2  & -4/9 & -4/9 & -4/9 \\
\hline
$A_\delta(k)$ decay & & & & \\
$f = 0$ (hydro)   & $\times$ & $\times$ & $\times$ & $\times$\\
$f = 10^{-3}$ & 100 kpc & 60 kpc & 45 kpc & 60 kpc \\ 
$f = 10^{-2}$ & $\sim L$  & 330 kpc & 240 kpc & $\sim L'=L/2$\\ 
$f = 10^{-1}$ & $>L\tablefootmark{{\it b}}$ & $>L\tablefootmark{{\it b}}$ & $>L\tablefootmark{{\it b}}$ & $>L'\tablefootmark{{\it b}}$\\
$f = 1$ & $>L\tablefootmark{{\it b}}$ & $>L\tablefootmark{{\it b}}$ & $>L\tablefootmark{{\it b}}$ & $>L'\tablefootmark{{\it b}}$\\
\hline\hline
\end{tabular}
\tablefoot{
\tablefoottext{\it{a}}{The energy and power spectrum slopes are retrieved through $E(k)_\delta= A(k)_\delta^2\, k^{-1}$ and $P(k)_\delta\propto A(k)_\delta^2\, k^{-3}$. For instance, the $A(k)_\delta$ slopes -1/3 and -1/2 correspond to the classic Kolmogorov and Burgers energy slopes -5/3 and -2, commonly observed for the velocity energy spectrum.}
\tablefoottext{\it b}{Models with strong conduction ($f\gta0.1$) produce a suppression of $\delta$ perturbations over the whole range of scales, inducing a decrease in normalization. The $A(k)_\delta$ decay occurs near $P_{\rm t} \sim 100$. 
Notice that the decay is {\it not} a sharp cutoff, due to the continuous turbulent regeneration. This is characterized by an exponentially changing slope, especially in the models with $f\lta10^{-2}$ (denoted with the $\rightarrow$ symbol).} 
}
\end{table*}

We now describe the results of the simulated models, 
reminding that the main goal of the present investigation is to understand the role of turbulence {\it and} conduction in shaping the power spectrum of density perturbations, $A(k)_\delta$ (\S\ref{s:filter}; we defer to future work the study of other statistics, as PDFs and structure functions). We are interested in the characteristic level of $\delta\rho/\rho$ perturbations, the slope of the spectrum, as well as any evident decline (or cutoff). Table \ref{table:1} summarizes the key retrieved properties and serves as a guide for the analysis of $A(k)_\delta$.

A key quantity for describing the evolution of perturbations is the ratio of the conduction and turbulence timescale (Eq. \ref{e:tturb} and \ref{e:tcond}), which normalized to reference values of the unsuppressed conductive run results to be
\begin{equation} \label{t_turbcond}
\frac{t_{\rm cond}}{t_{\rm turb}} \simeq \,\,1 \,\: l_{550}^{4/3}\:\:\left[\frac{\sigma_{v,\rm 370}\,\, n_{\rm e, 0.004}}{L^{1/3}_{650}\,\,f_{1}\,\,T^{5/2}_{\rm e, 8}}\right].
\end{equation}
We do not consider saturated conduction for this estimate, since the temperature gradient is not steep for the majority of the zones (the interpolated $D_{\rm cond}$ shall be used in this case for higher accuracy; \S\ref{s:cond}).
This key timescale ratio can be also seen as the Prandtl number applied to turbulence, instead of to the kinematic viscosity, which we define as
\begin{equation} \label{Pt}
P_{\rm t} \equiv \frac{D_{\rm turb}} {D_{\rm cond}} = \frac{l^2/D_{\rm cond}}{l^2/D_{\rm turb}} = \frac{t_{\rm cond}}{t_{\rm turb}},
\end{equation}
where the thermal and turbulent diffusivities are provided in \S\ref{s:turb}$\,$-$\,$\ref{s:cond}. The reference value is usually taken at the injection scale, $l=L$.
Remarkably, the qualitative evolution of a very complex nonlinear dynamics can be approximately predicted via the dominant timescale ratio (or dimensionless number; see discussion in \S\ref{s:time}). For instance, different values of $f$ and $M$ can lead to the same $P_{\rm t}$, hence to a similar qualitative dynamics and power spectrum of density perturbations (cf. \S\ref{r:med} and \ref{r:weak_half}).

\subsection{Weak turbulence: $M\sim 0.25$} \label{r:weak}

\begin{figure} 
    \begin{center}
       \subfigure{\includegraphics[scale=0.42]{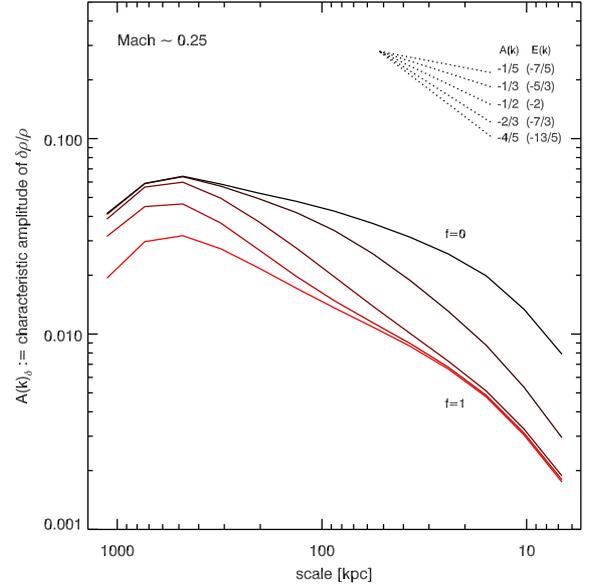}}
       \caption{Characteristic amplitude of $\delta \rho/\rho$, $A(k)_\delta=\sqrt{P(k)_\delta\,4\pi k^3}$, for the models with weak turbulence $M\sim0.25$ and varying conduction, after reaching statistical steady state ($\sim$$2\ t_{\rm eddy}$)
       with the same level of continuous stirring. The driving is initiated only above 550 kpc.
       From top to bottom curve (black to bright red), the suppression of conduction is $f=0,\, 10^{-3},\, 10^{-2},\, 10^{-1},\, 1$. First column of Table \ref{table:1} summarizes the key properties. 
       Strong conduction globally damps $\delta$ perturbations by a factor of 2$\,$-$\,$4, substantially steepening the 
       spectrum and departing from the Kolmogorov slope of the no-conduction run. Weak conduction is able to induce the steep decay only near the scale linked to $P_{\rm t}\sim 100$.}
       \label{fig:Ak1}
     \end{center}
\end{figure}

The first set of models implements a low level of stirring, with average mass-weighted 3D Mach number $M\,$$\sim\,$0.25 ($\sim$$370\ {\rm km\ s^{-1}}$). Observations and simulations suggest in fact that turbulence in the ICM typically remains subsonic (\citealt{Nagai:2007, Nagai:2013, Vazza:2011, Gaspari:2012b, Sanders:2013}). The turbulent energy is $\sim$3.5 percent of the total thermal energy, $E_{\rm turb} \simeq 0.5\, \gamma (\gamma-1)\, M^2 E_{\rm th}\simeq 0.56\, M^2 E_{\rm th}$. 
Dissipational heating is thus negligible. In the current models, $t_{\rm eddy}=L/\sigma_v\sim1.6$ Gyr.
We always analyze the system as soon as it establishes statistical steady state, i.e.~after $\sim$$2\, t_{\rm eddy}$.

\begin{figure} 
    \begin{center}
       \subfigure{\includegraphics[scale=0.54]{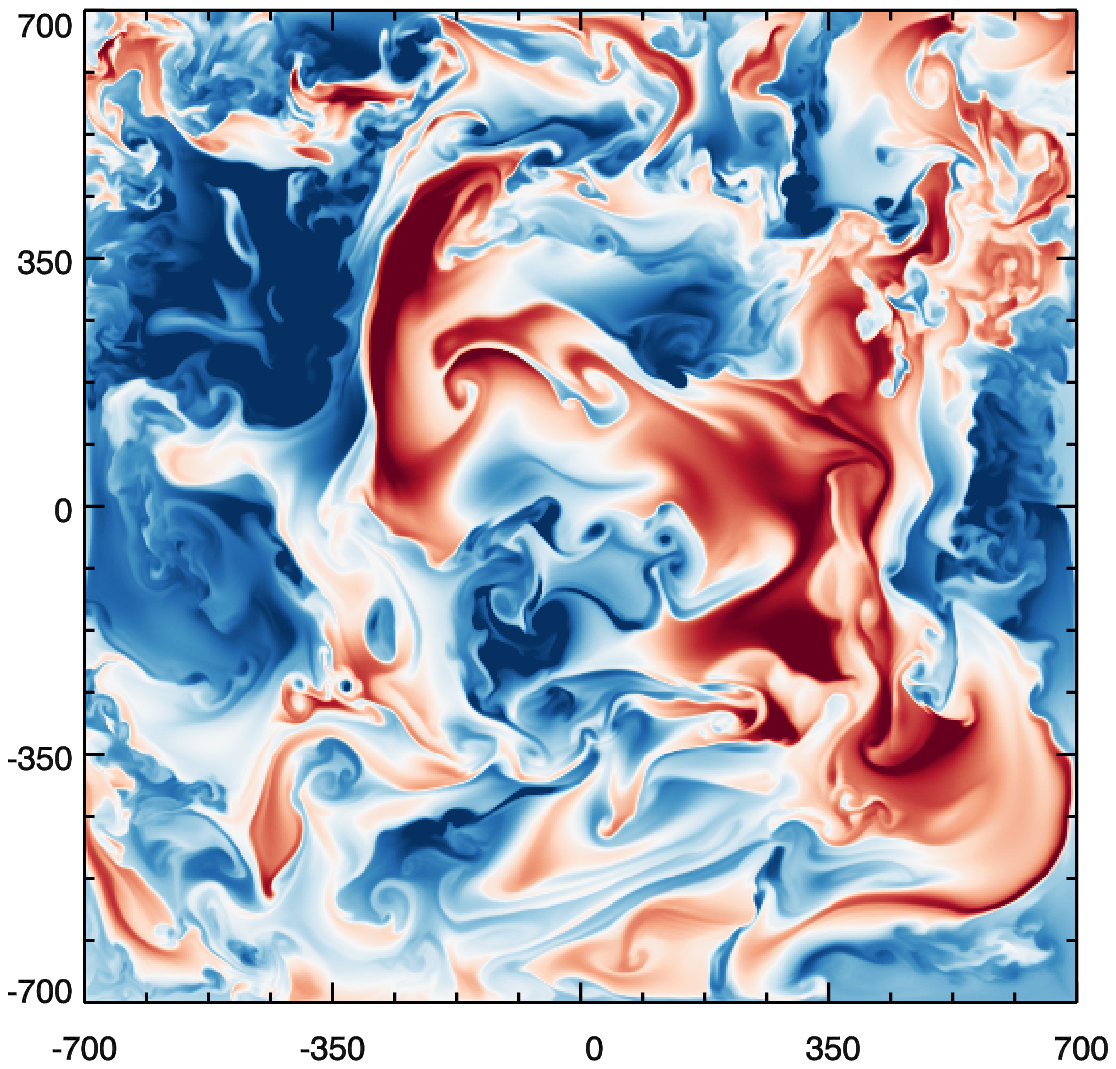}}
       \subfigure{\includegraphics[scale=0.54]{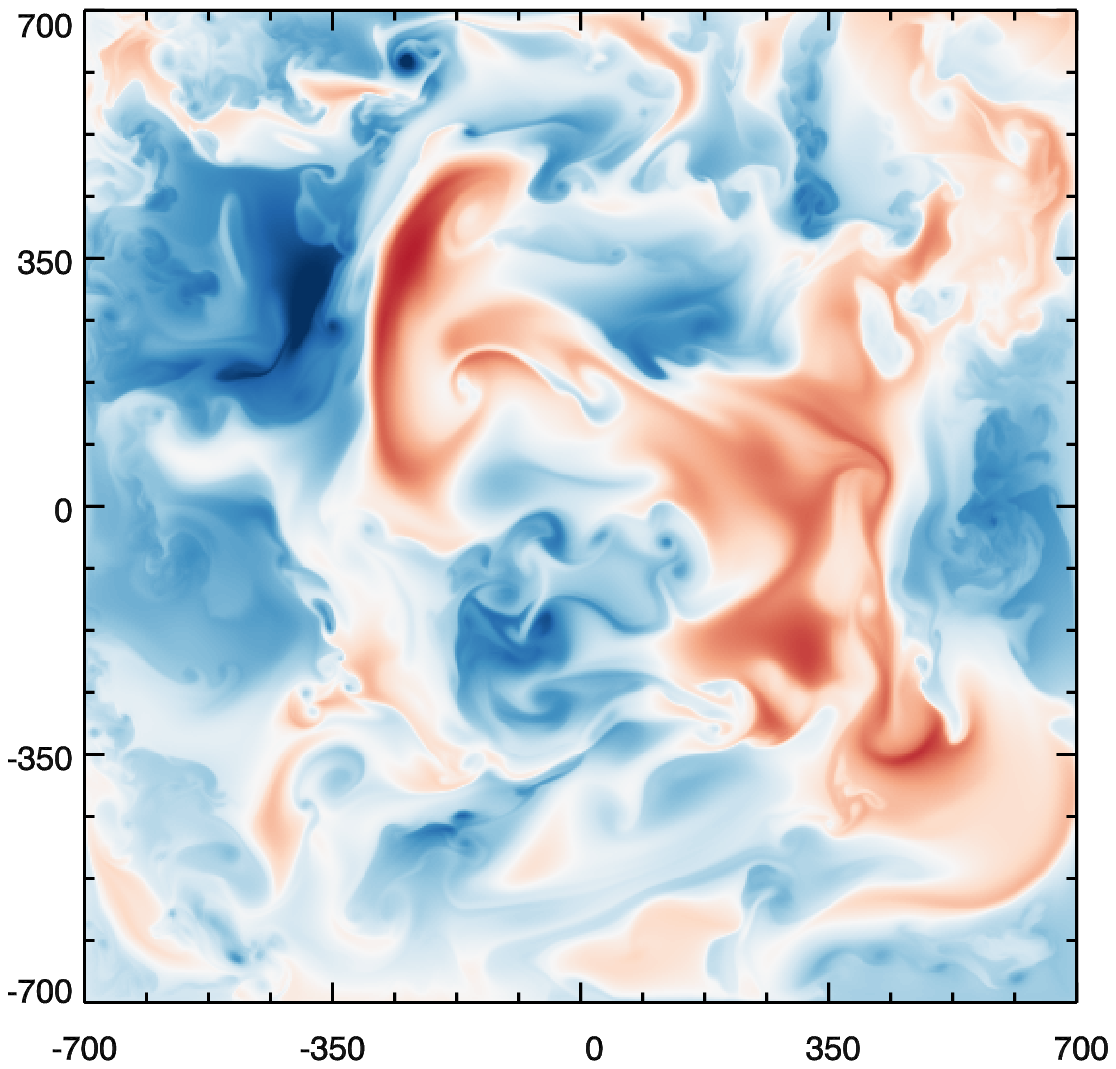}}
       \subfigure{\includegraphics[scale=0.54]{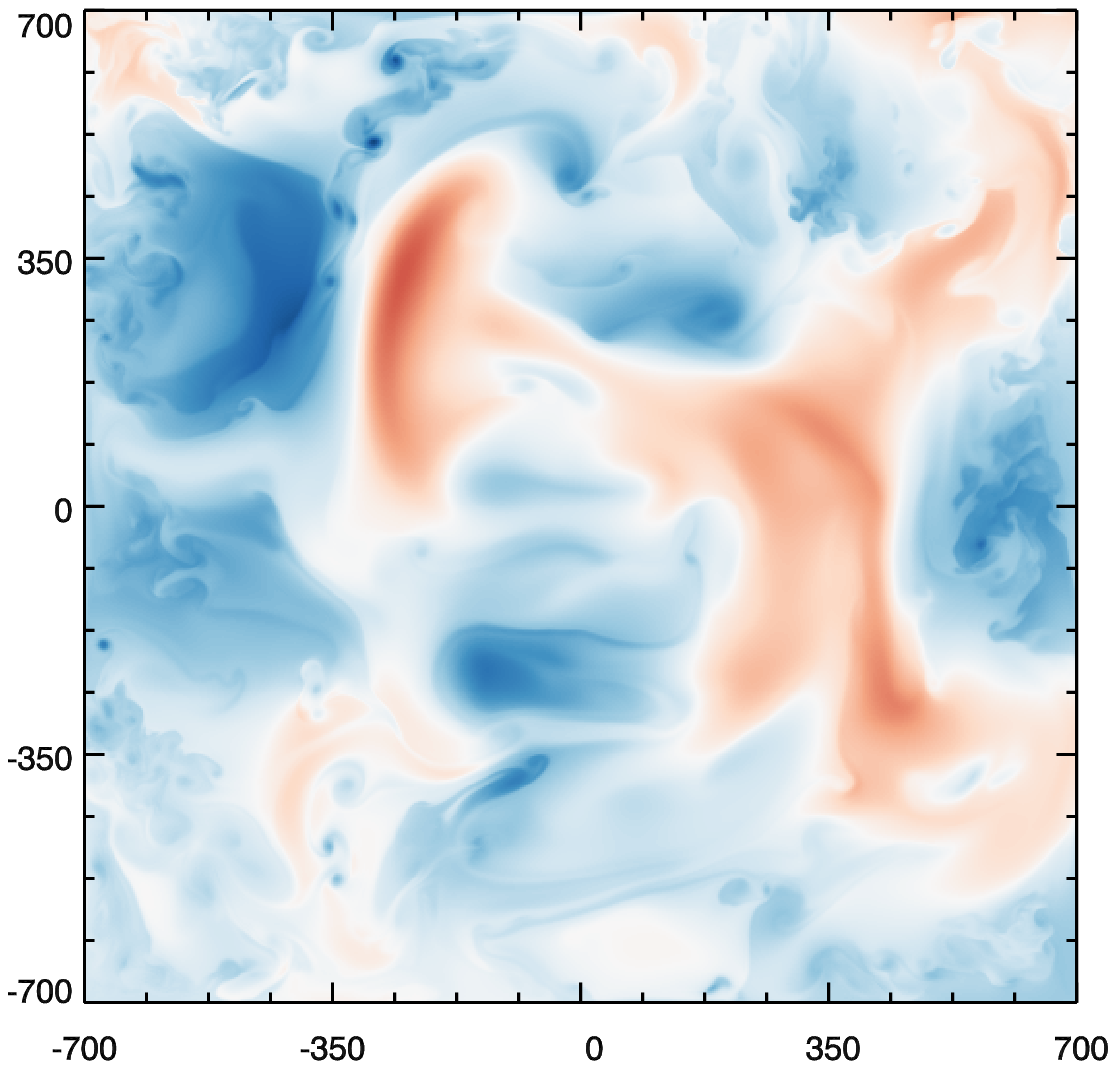}}
       \subfigure{\includegraphics[scale=0.54]{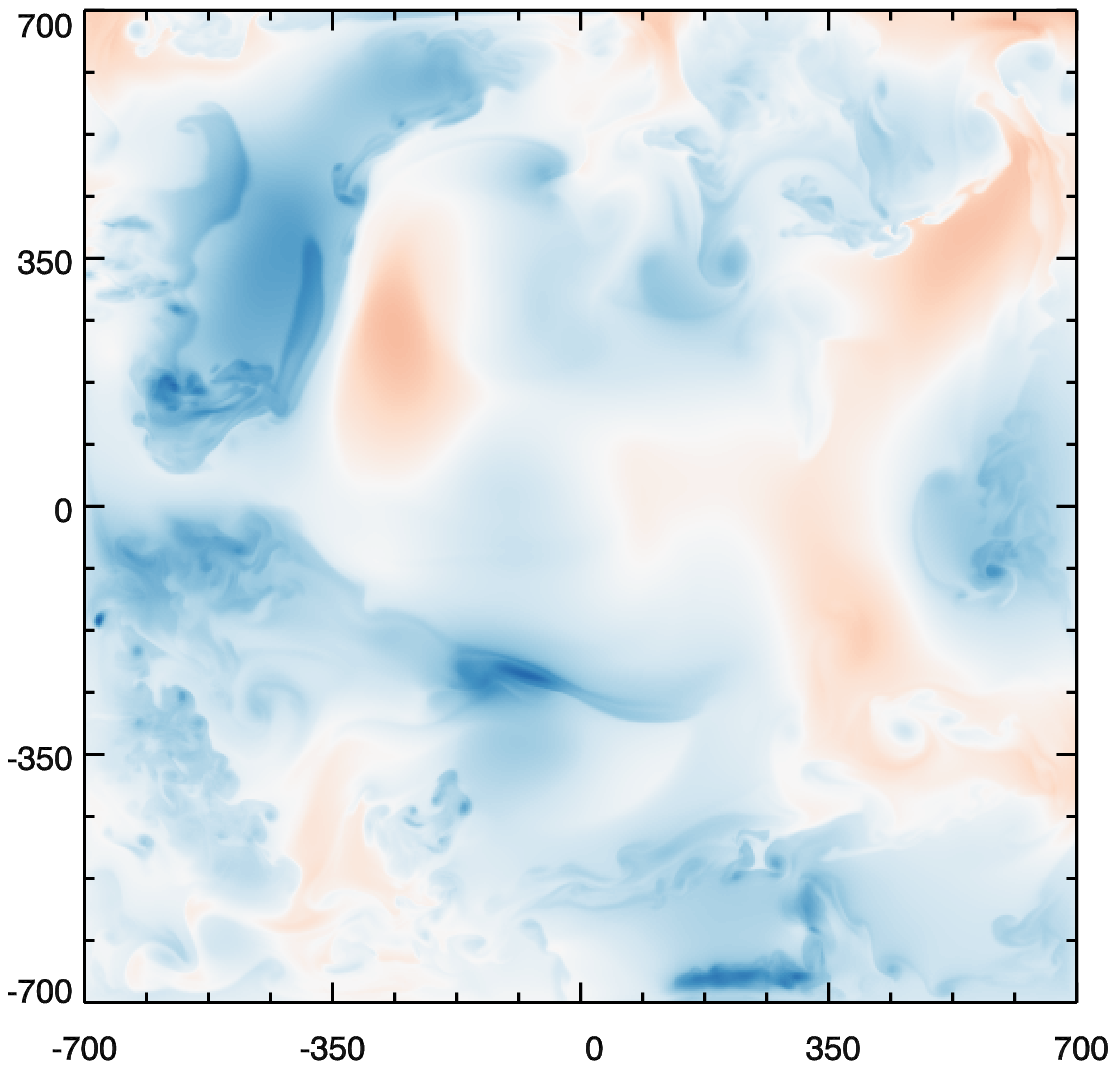}}
       \caption{Mid-plane cuts of $\delta \rho/\rho$ for the models with $M\sim0.25$. From top to bottom: $f=0,\, 10^{-3},\, 10^{-2},\, 10^{-1}$ (the latter very similar to $f=1$ run). The color coding is blue $\rightarrow$ white $\rightarrow$ red: -40\% $\rightarrow$ 0\% $\rightarrow$ 40\%. 
     \label{fig:delta}}
     \end{center}
\end{figure}

\begin{figure*} [!pth] 
       \centering
       \subfigure{\includegraphics[scale=0.47]{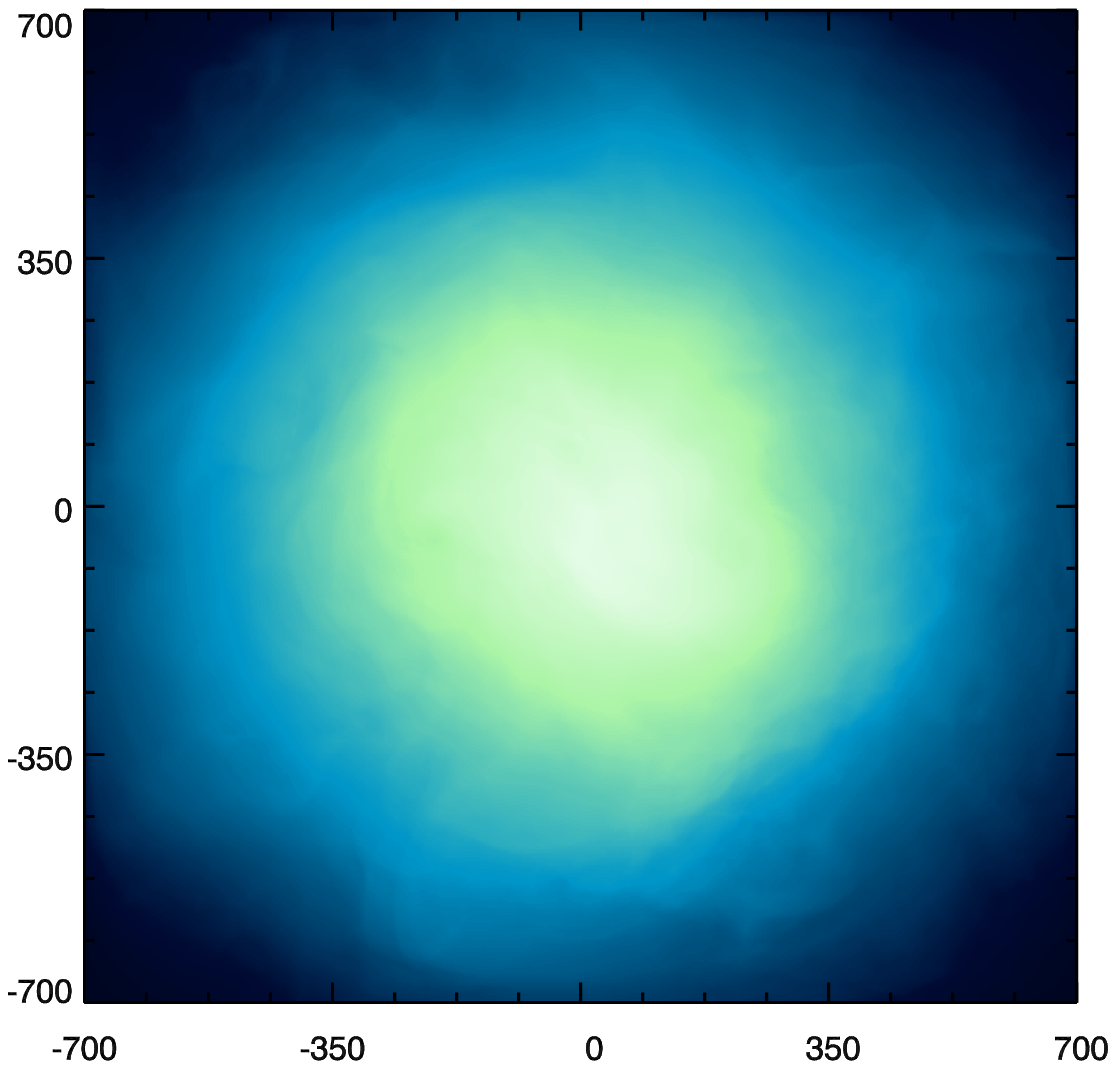}}
       \subfigure{\includegraphics[scale=0.47]{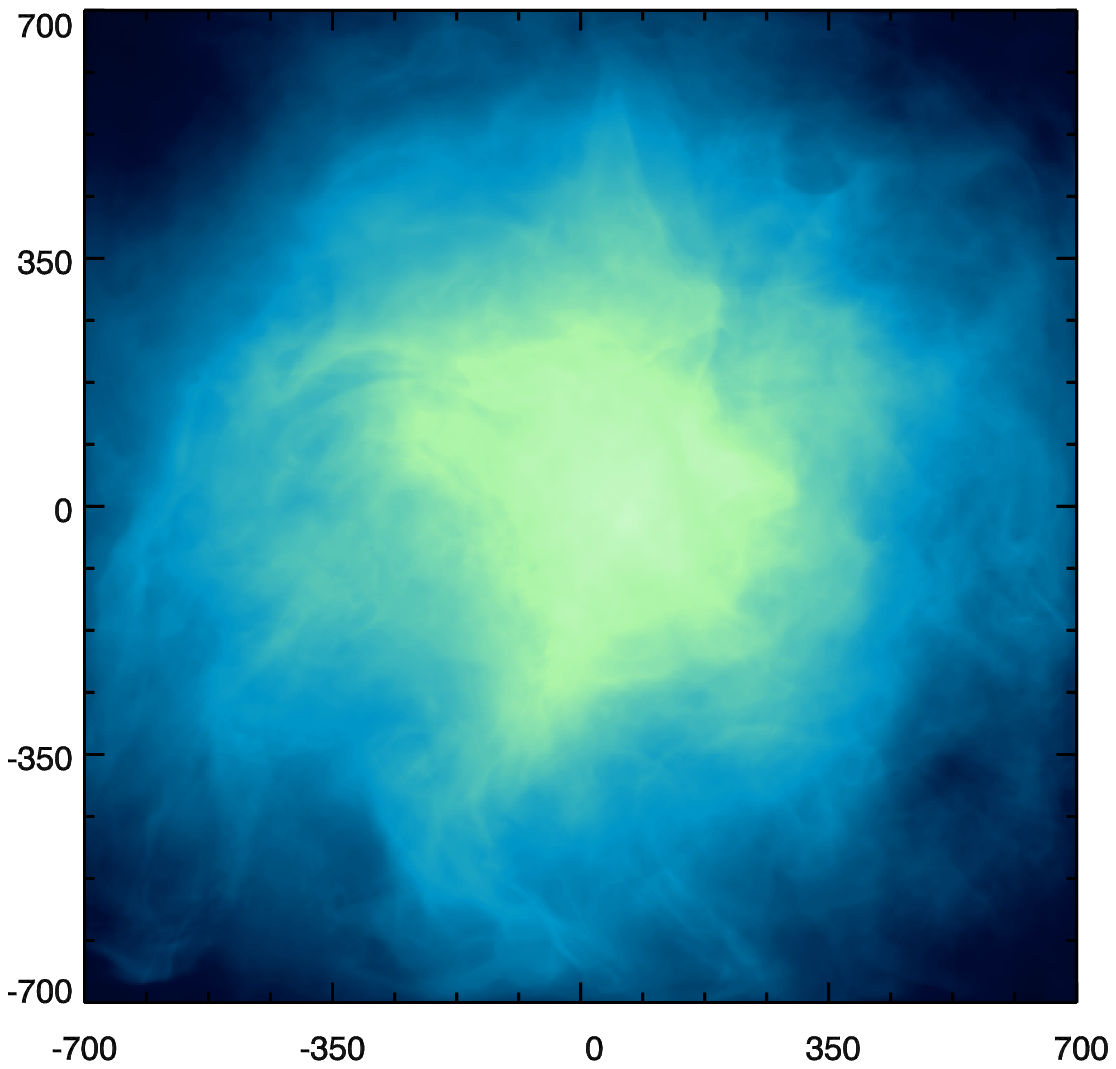}}
       \subfigure{\includegraphics[scale=0.47]{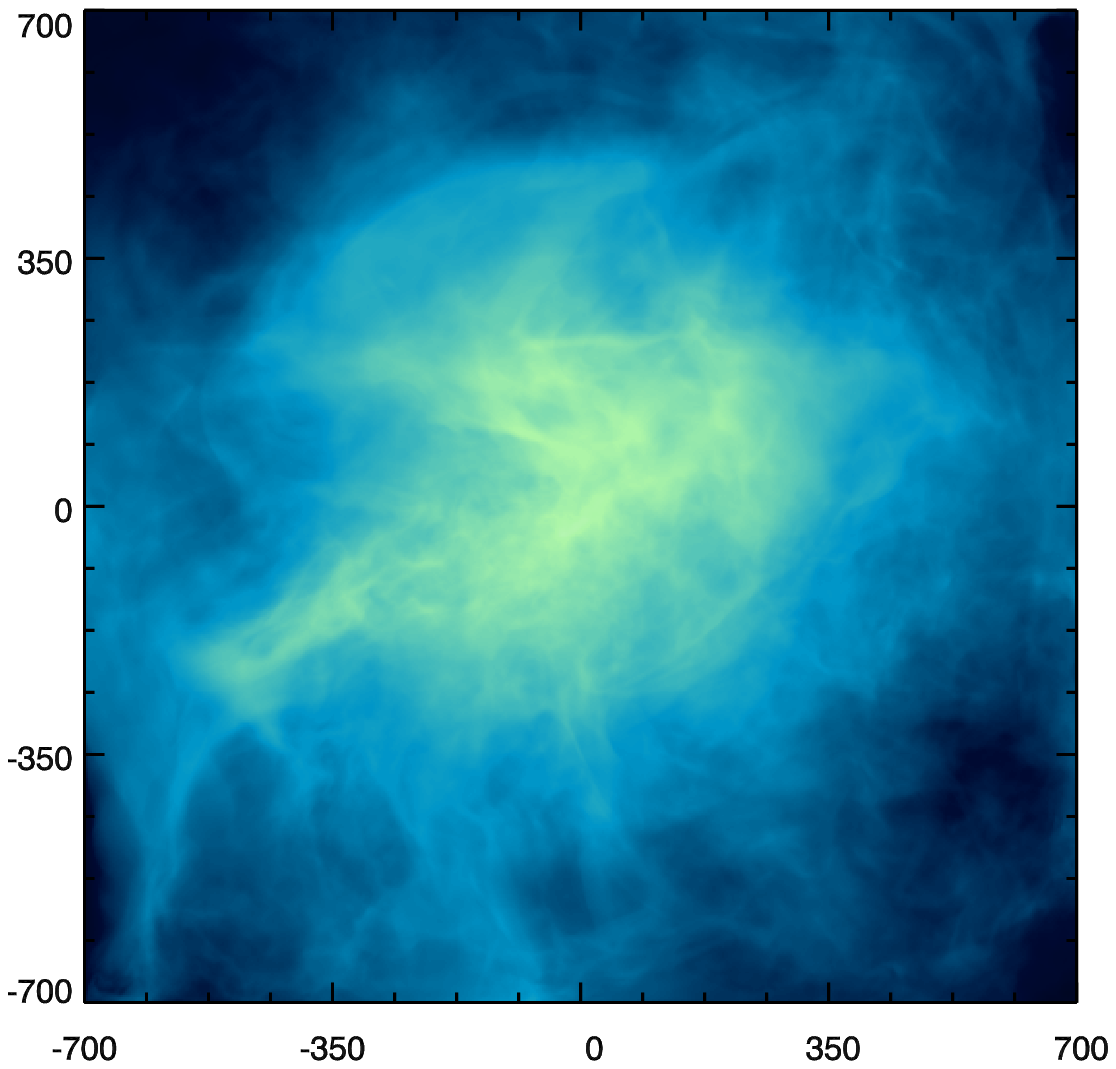}}       
       \subfigure{\includegraphics[scale=0.47]{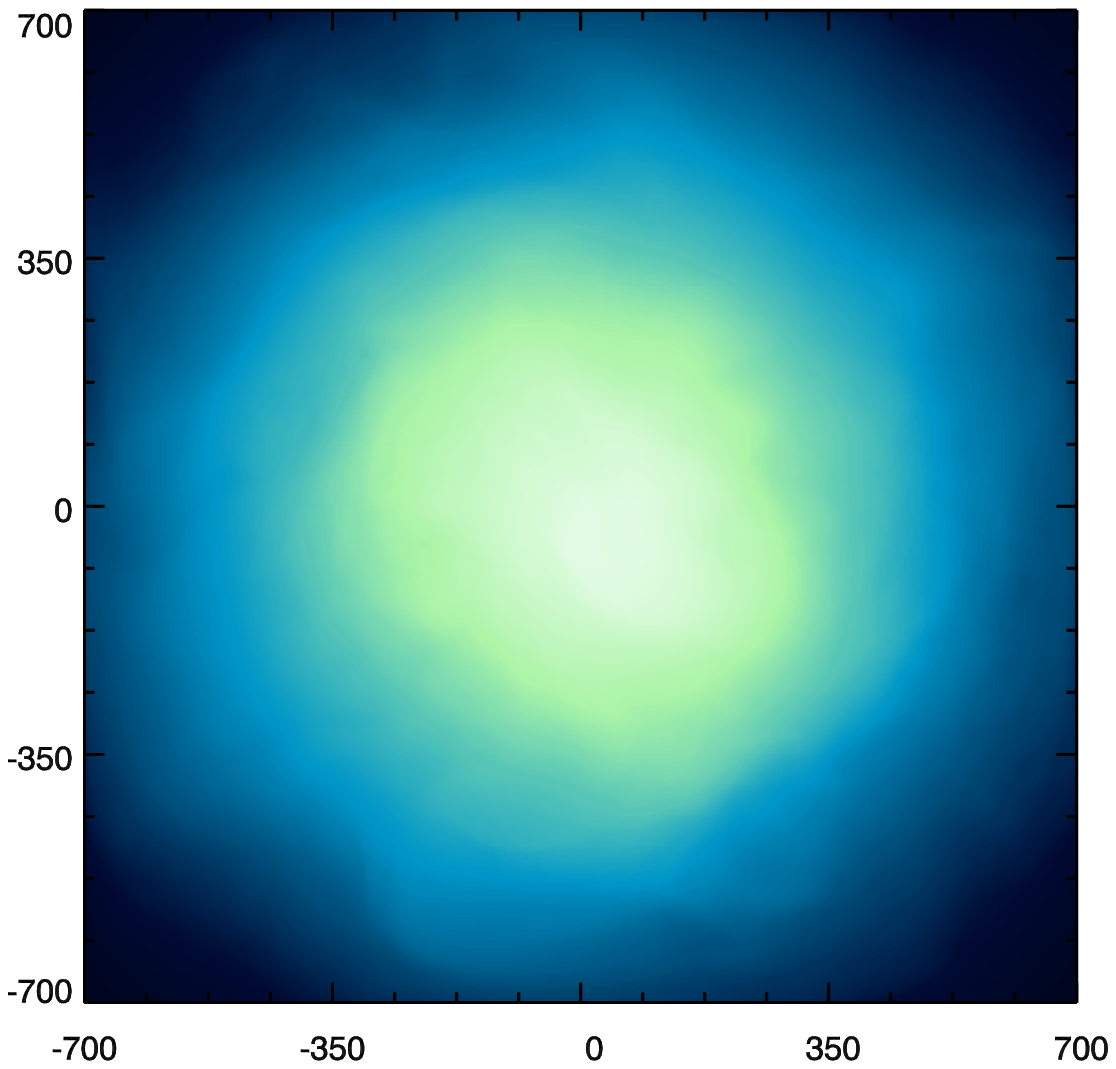}}
       \subfigure{\includegraphics[scale=0.47]{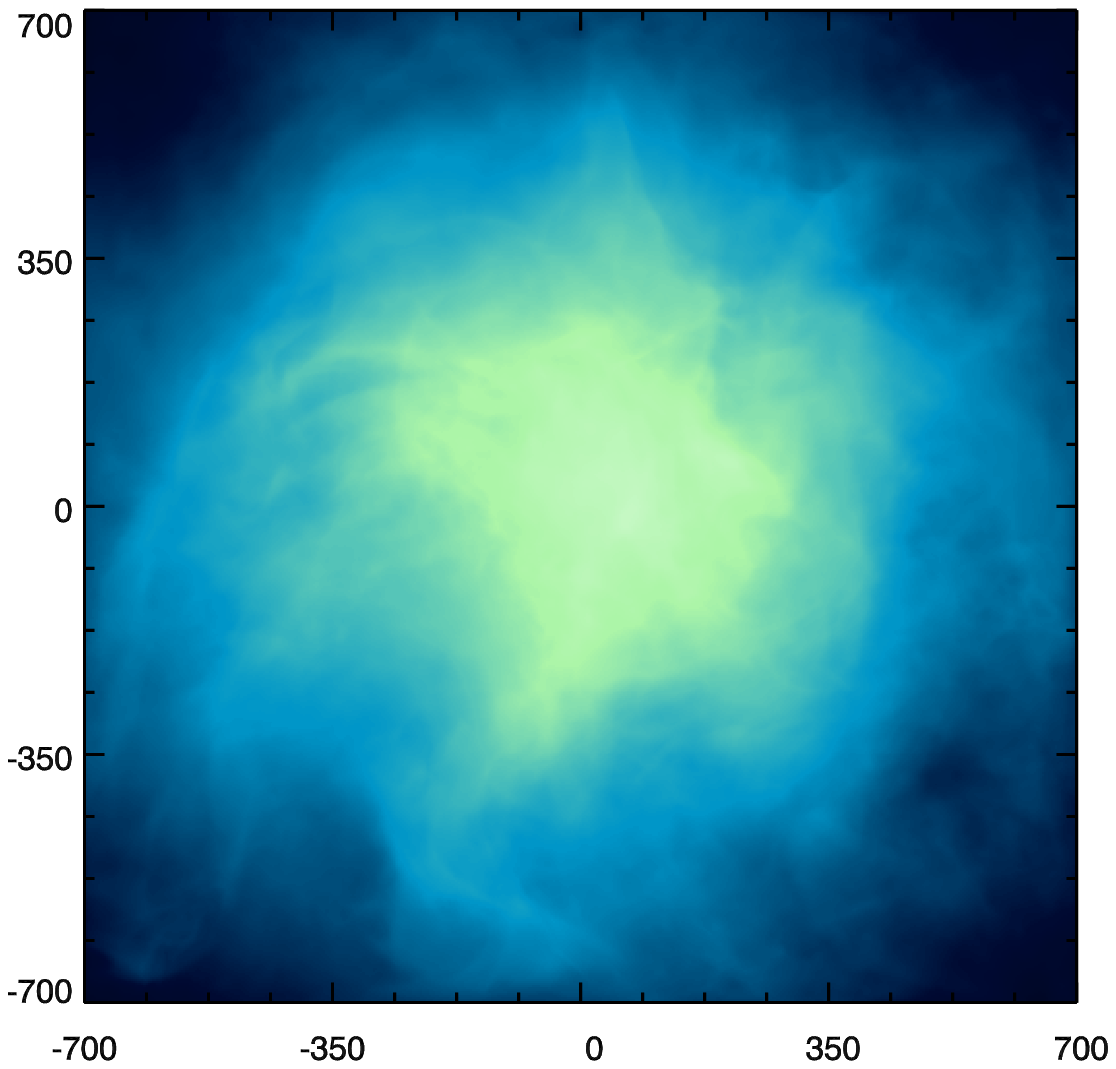}}
       \subfigure{\includegraphics[scale=0.47]{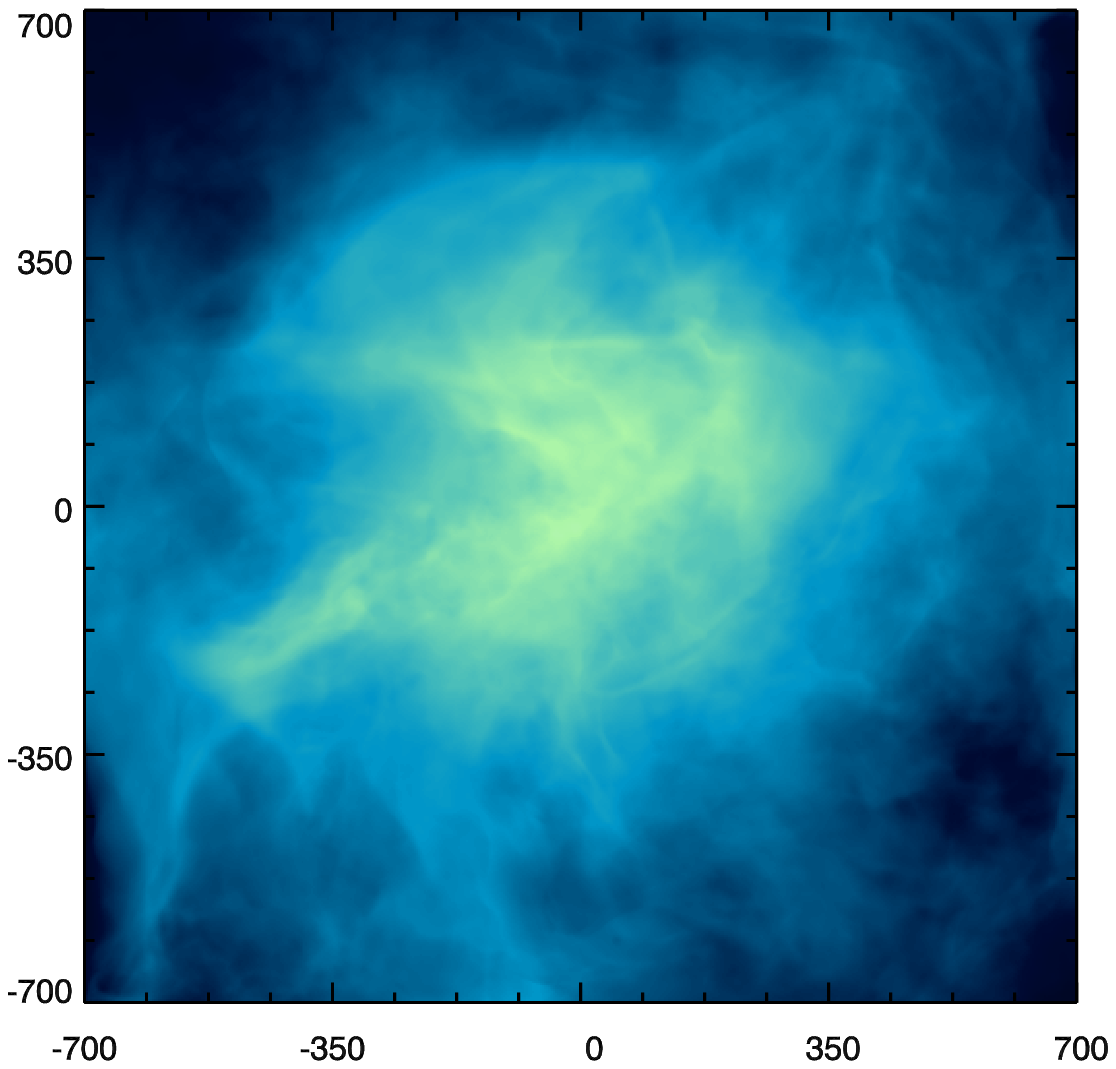}}       
       \subfigure{\includegraphics[scale=0.47]{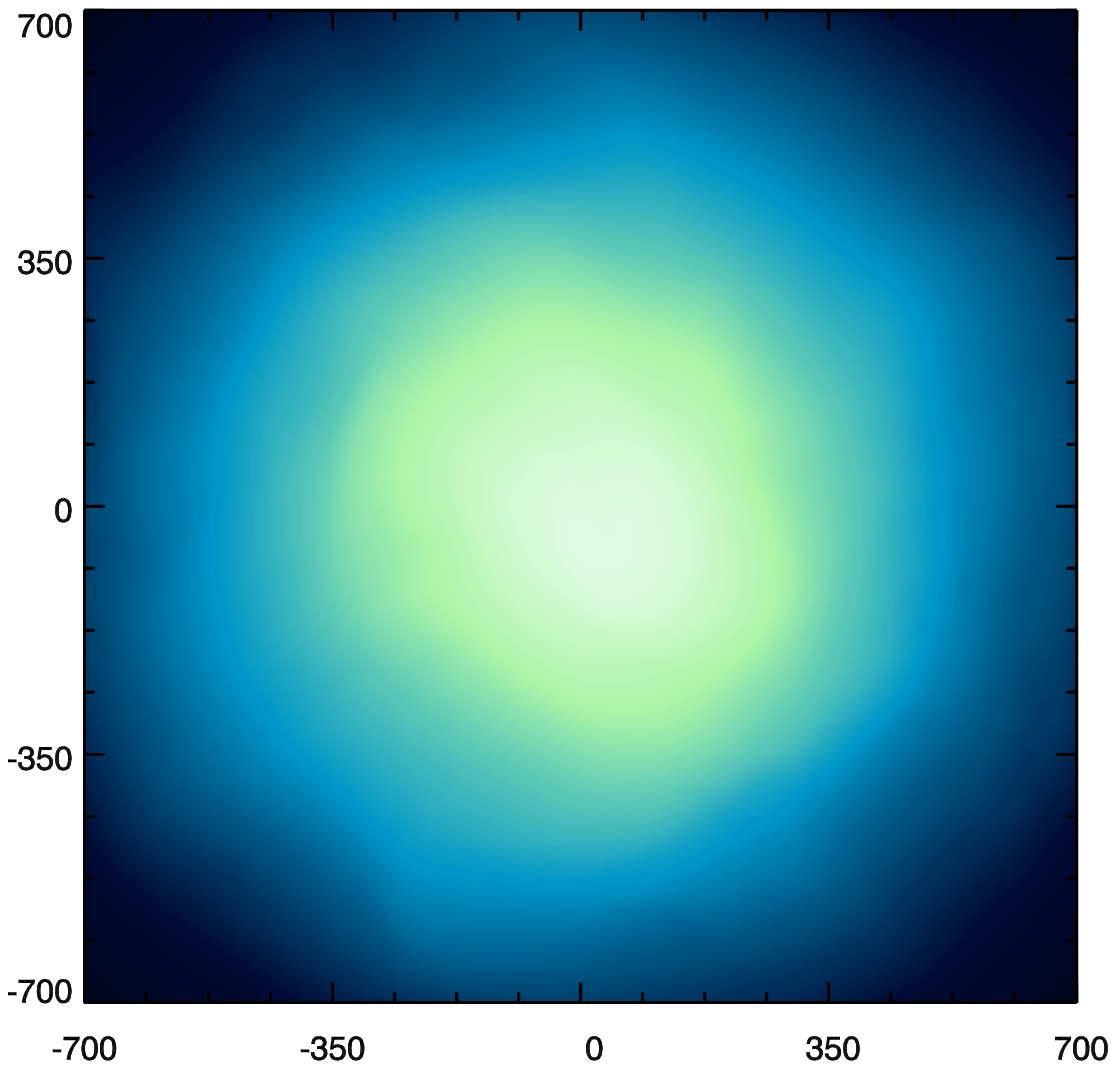}}
       \subfigure{\includegraphics[scale=0.47]{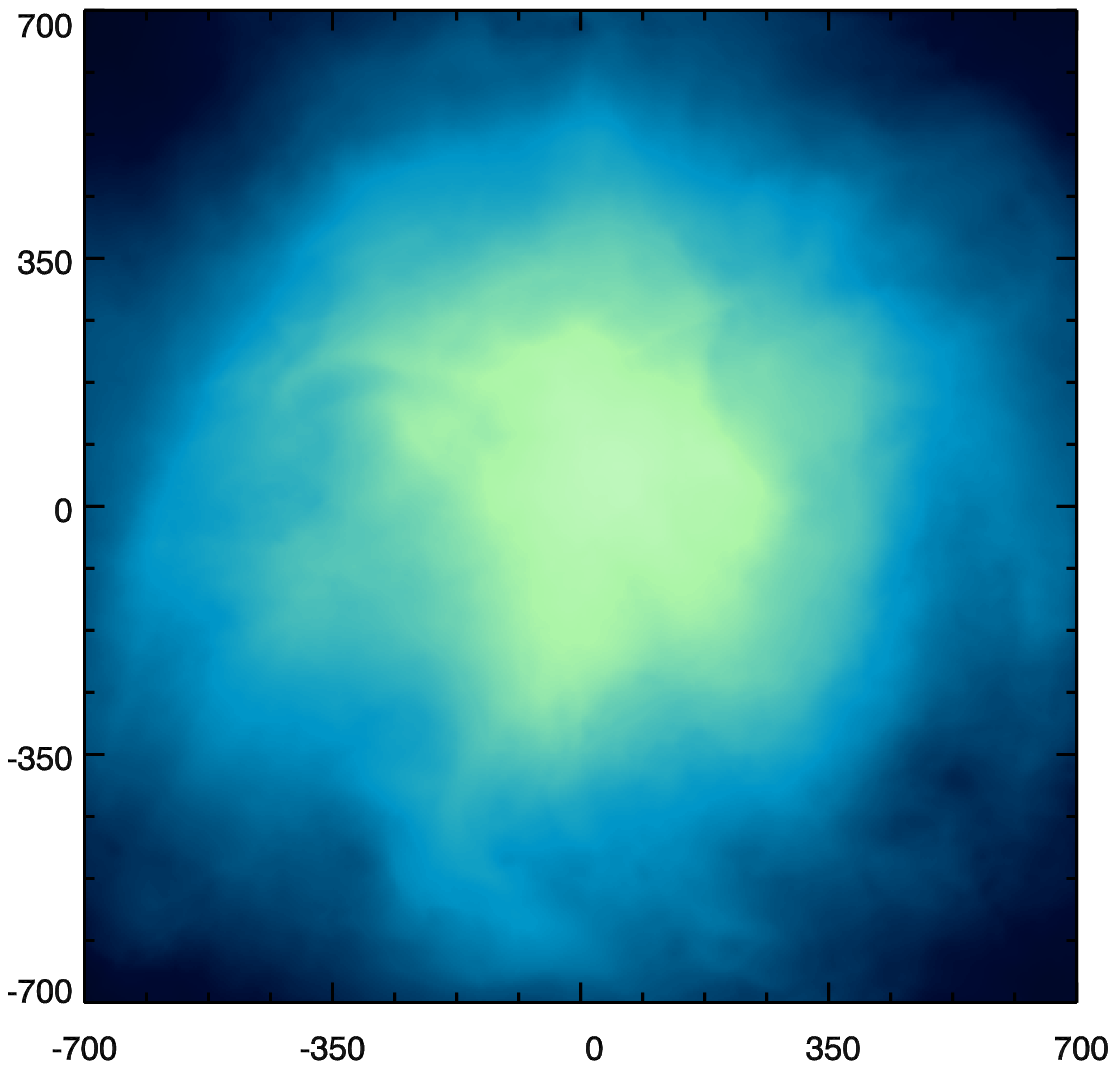}}
       \subfigure{\includegraphics[scale=0.47]{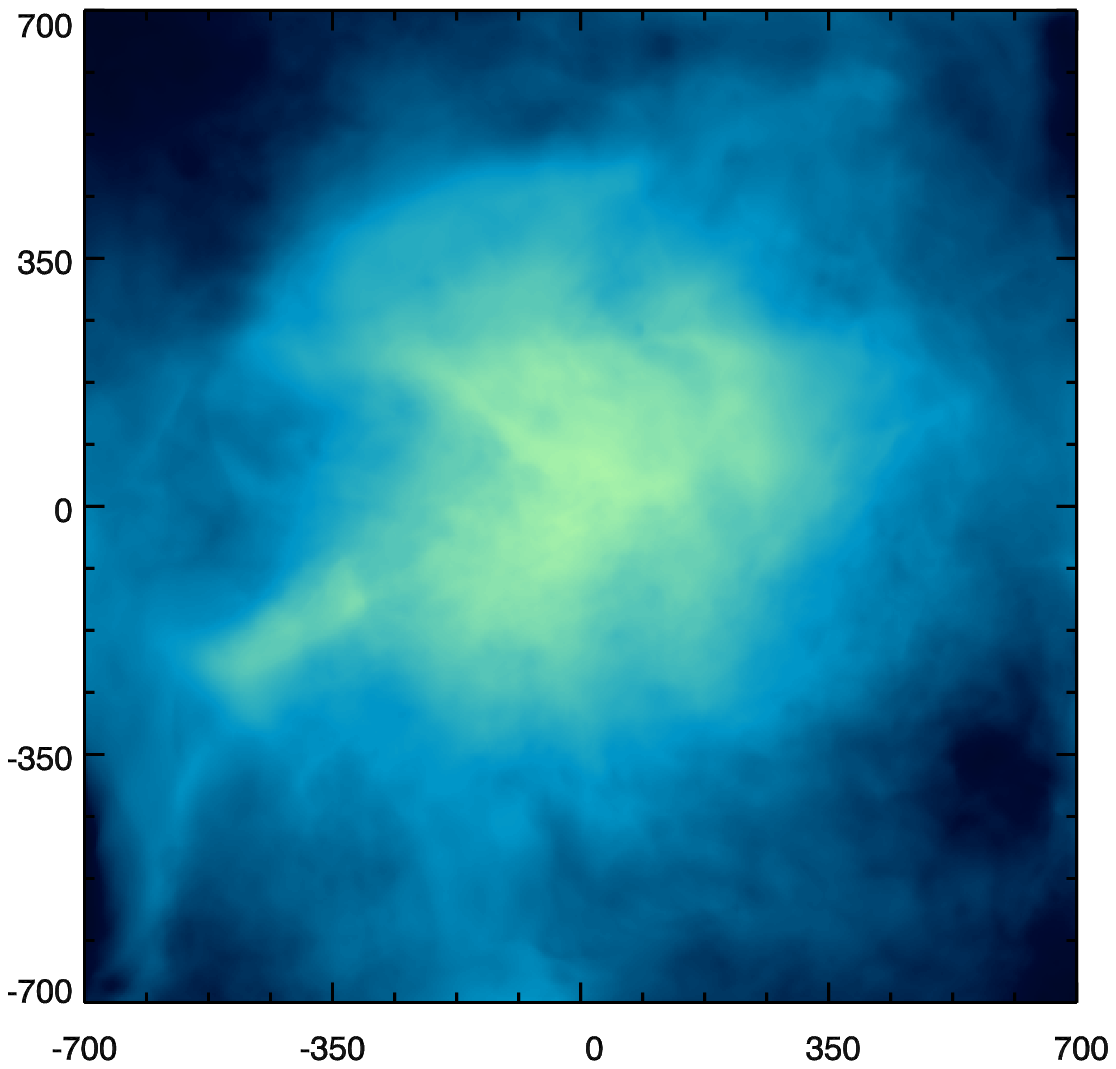}}       
       \subfigure{\includegraphics[scale=0.47]{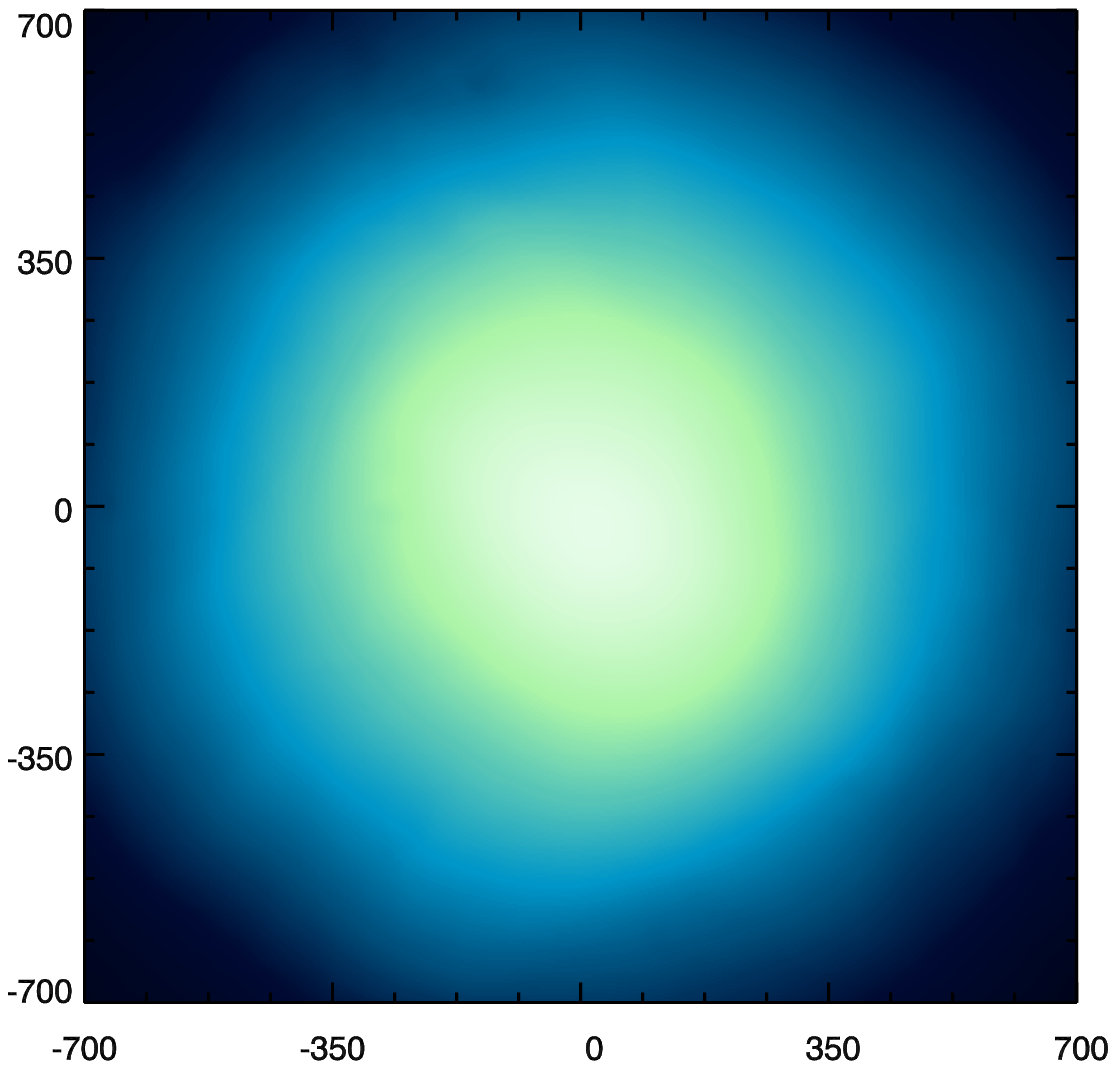}}
       \subfigure{\includegraphics[scale=0.47]{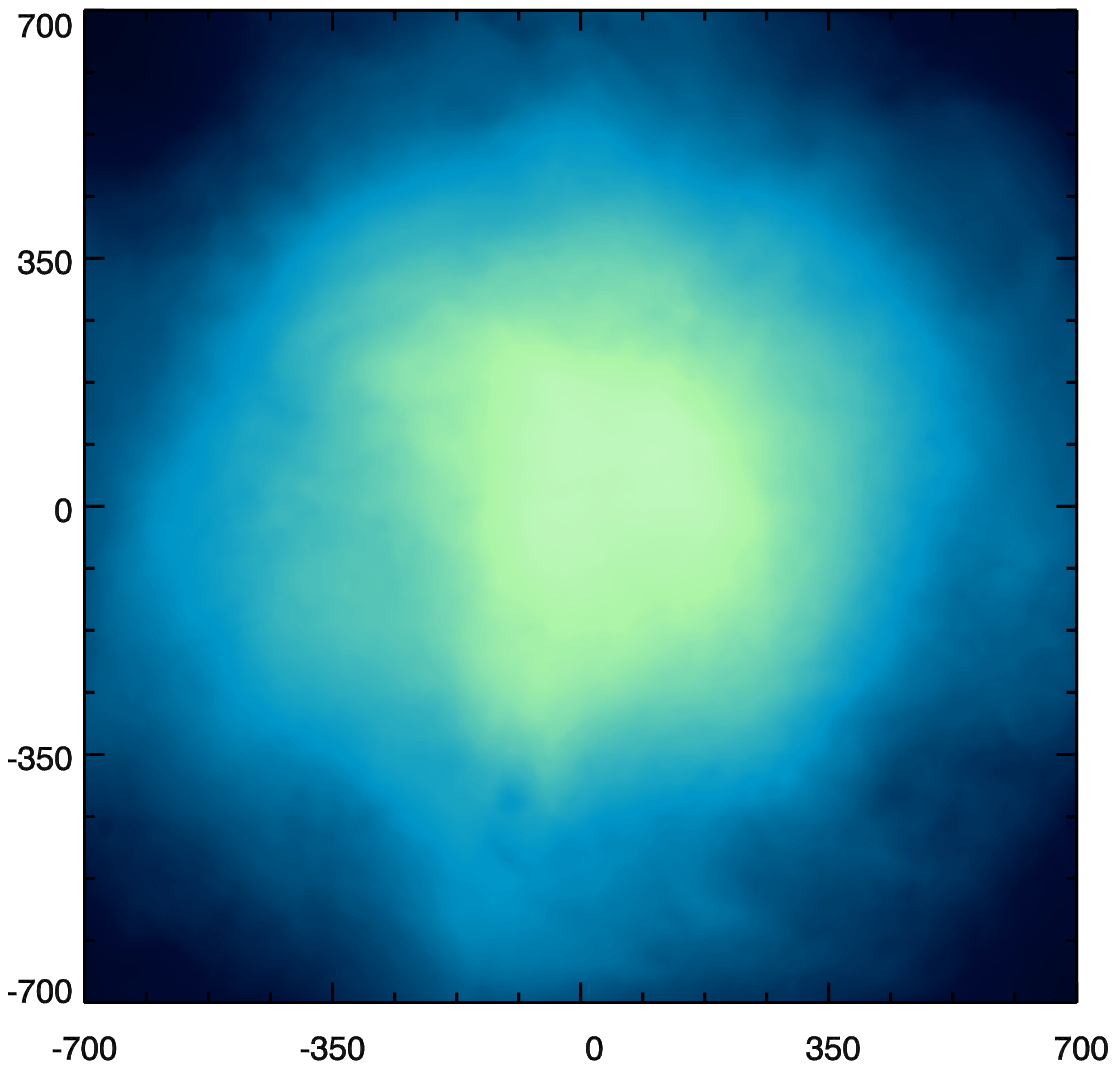}}
       \subfigure{\includegraphics[scale=0.47]{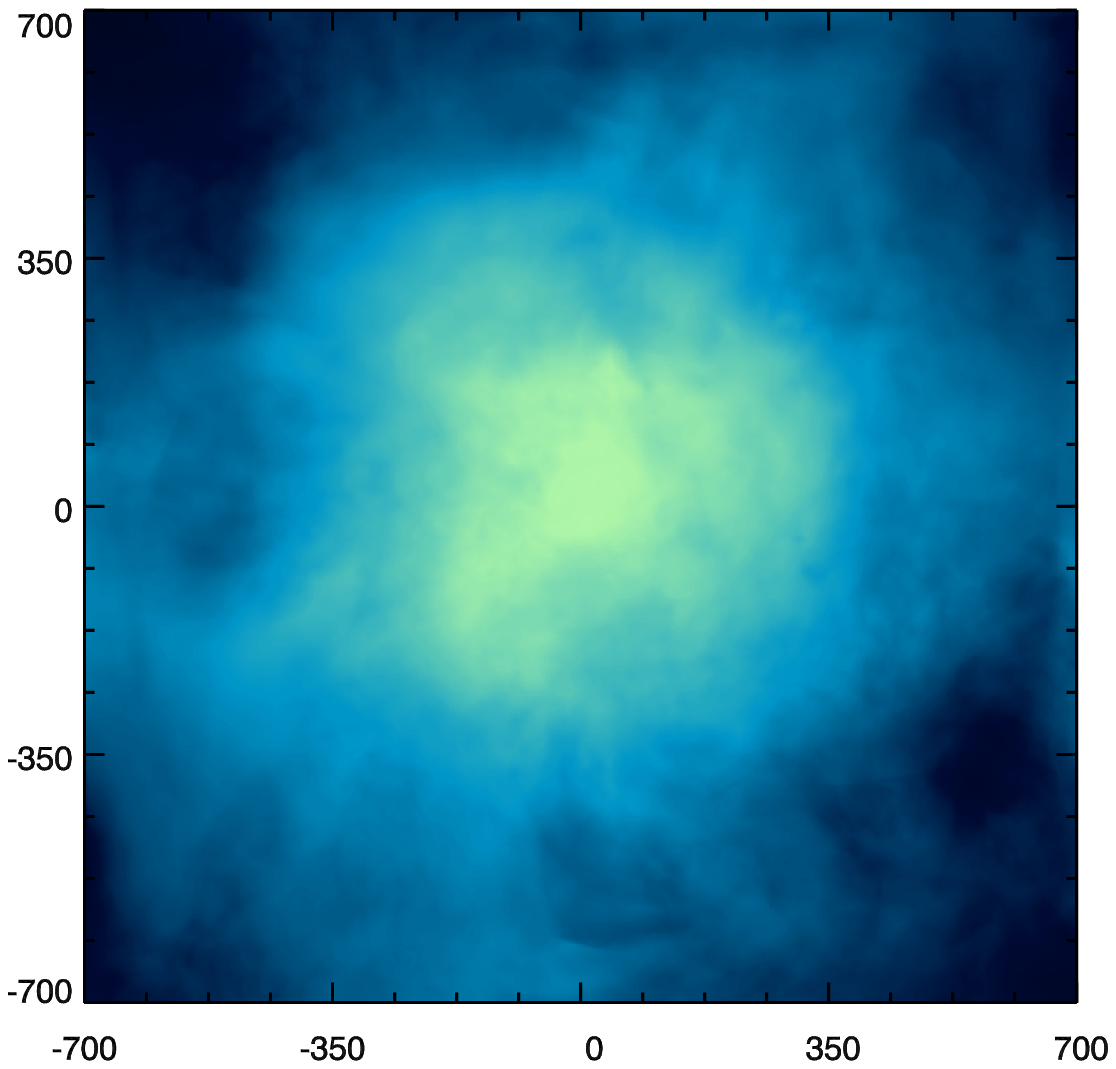}}             
       \caption{X-ray surface brightness for the models with weak ($M\sim0.25$; left), mild ($M\sim0.5$; middle), and strong
       turbulence ($M\sim0.75$; right). From top to bottom: $f=0, \,10^{-3}, \,10^{-2}, \,10^{-1}$ ($f=1$ maps are similar to the latter images). 
       The color coding is black - blue - pale green, in the range $10^{-7}$ - $4\times10^{-5}$ erg s$^{-1}$ cm$^{-2}$.
       Conduction prevents the development of the full turbulent cascade, especially if $f\ge0.1$. K-H and R-T rolls and filaments are thus suppressed, and the cluster retains the spherical, smooth shape. Strong turbulence is instead able to deform the cluster, flattening the entropy profile and inducing a fainter (more rarefied) core.
       Perturbations are best observed at $r> r_{\rm c}$, or over the whole cluster if $M>0.5$.} 
     \label{fig:SBx}
\end{figure*}

In the purely hydrodynamic run ($f=0$, i.e.~no conduction or \Pt$\rightarrow\infty$) the driven stirring motions generate a
turbulent cascade in the $\delta\rho/\rho$ power spectrum analogous to that of the turbulent velocities. The characteristic amplitude shows the typical injection peak at low $k$ ($l\sim600$ kpc), followed by the inertial range and the final steepening due to dissipation (Fig.~\ref{fig:Ak1}, black line).
The characteristic level of perturbations, given by the maximum of $A(k)_\delta$, is 6.5 percent (6.7 percent using FFTs).
An important result is that the inertial range of the density perturbations is remarkably similar to the Kolmogorov slope, $A(k)_\delta\propto k^{-1/3}$ (or $E(k)\propto k^{-5/3}$), slightly flattening towards the injection scale. Stratification has overall a secondary impact on $\delta$ (\S\ref{s:time}). 
Dissipation via numerical viscosity (mimicking Spitzer viscosity; \S\ref{s:visc}) becomes substantial below 6 resolution elements. 
Notice that current observations are limited to scales $\gta\,$30 kpc due to Poisson noise and projection (Fig.~\ref{fig:comp}), which are well resolved by the current runs.  
The dynamics is driven by key hydrodynamical instabilities, as Kelvin-Helmholtz (K-H) and Rayleigh-Taylor (R-T), inducing the characteristic rolls, curls, and edges in both $\delta$ (Fig.~\ref{fig:delta}) and SB$_{\rm x}$ maps (Fig.~\ref{fig:SBx}, top).

Overdense and underdense regions are associated with a mixture of isobaric and adiabatic perturbations, which we analyze in \S\ref{s:modes}. With weak stirring the former dominates, while strong turbulence enhances the adiabatic mode (analogous to pressure waves).
Unlike in the strongly conductive runs, the entropy gradient becomes progressively shallower, inducing a lower central density (30 percent) and higher temperature (10 percent). 
The transport of heat due to turbulent diffusion is $\propto D_{\rm turb}\boldsymbol{\nabla} s$ (Eq.~\ref{e:turb_heat}). Therefore, turbulence seeds perturbations in density and temperature, while diffusing entropy. Conduction diffuses instead temperature {\it and} density fluctuations.

In the next experiment, we enable electron thermal conduction. 
We start analyzing the unmagnetized case ($f=1$), though even in the unsuppressed regime, conduction can be limited by the saturated flux (\S\ref{s:cond}).
The overall dynamics and $A(k)_\delta$ is however not affected by saturation; the fraction of saturated zones is less than 10 percent (becoming $\lta1$ in the $f=0.1$ run). 
The discrepancy between the purely turbulent and the conductive run is evident in the density perturbations (Fig.~\ref{fig:Ak1}, red line), allowing to put critical constraints on the physical properties of the ICM. Three are the key modifications imparted by conduction. First, density perturbations are significantly damped over the whole range, by a factor of 4 on small scales to a factor of 2 on large scales, where the peak of perturbations reaches 3.5 percent.
Second, the slope after the injection hump is considerably steeper than Kolmogorov, 
following a Burgers-like\footnote{`Burgers' name is just used as slope reference; the steepening is here not produced by shocks as in classic supersonic turbulence.} spectrum $k^{-1/2}$ ($E(k)\propto k^{-2}$). Third, there is no evident cutoff, meaning that conduction efficiently operates on all scales, while 
turbulence is not able to consistently regenerate perturbations. Considering the dimensionless turbulent Prandtl number (Eq. \ref{Pt}), at the injection scale \Pt$\sim1$ (Fig.~\ref{fig:time1}). Albeit thermal diffusion is a small-scale process (\Pt $\propto l$), it is ubiquitous and quick enough to efficiently stifle the full turbulent cascade in the whole cluster. 
The spectrum does not sharply decline to zero value, since turbulent regeneration is continuous. This is in analogy with the observed radio spectra, where relativistic electrons are regenerated by turbulence, preventing a dramatic cutoff.

The $\delta \rho/\rho$ maps of the strongly conductive runs (Fig.~\ref{fig:delta}, bottom) 
clearly show the absence of significant density perturbations, especially on small scales. 
Since turbulent diffusion is severely inhibited, the cluster strongly retains the initial spherical symmetry and radial profiles, as indicated by the X-ray surface brightness map in Figure \ref{fig:SBx} (bottom row). 
Only the electron temperature is able to become quickly isothermal both locally and globally; due to the non-negligible ion-electron equilibration delay, $\sim$50 Myr (\S\ref{s:equip}), the ion temperature has instead difficulty in becoming fully isothermal as $T_{\rm e}$. The discrepancy is in the range ($T_{\rm e}-T_{\rm i})/T_{\rm i}\sim1-15$ percent, from the inner to external radial shells (in particular beyond $r_{\rm c}$, where the decreasing densities increase the lagging; cf.~\S\ref{s:time}).
In the opposite regime, the purely turbulent run shows numerous filaments and depressions in density. In the SB$_{\rm x}$ image (top row), the perturbations are partially veiled by the line of sight integration; the ideal location to observe perturbations is at $r\gta r_{\rm c}$ (see Appendix \ref{app:2}).

We apply now a suppression factor of 10 on thermal conduction.
The regime $f\sim0.1-0.3$ is widely adopted in astrophysical
studies (e.g.~\citealt{Narayan:2001, Voigt:2004, Dennis:2005, Ruszkowski:2010, Ruszkowski:2011}; see \S\ref{s:MHD}). 
As shown in Figure \ref{fig:Ak1}, the characteristic amplitude is similar to the unsuppressed case.
This is a key result, telling us that density perturbations are suppressed regardless of commonly adopted suppression factors ($f\gta0.1$).
Observations could hence put strong limits on the suppression $f$, based on the steepness and/or normalization
of $A(k)_\delta$ spectrum (\S\ref{s:disc}).
Compared with $f=1$ model, the density fluctuations increase by a factor of 0.3 on large scales, while small perturbations have similar power. The absence of a dramatic decline is mainly due to the fact that on small scales the sound crossing time becomes greater than the conduction time (e.g.~Fig.~\ref{fig:time1}), hence the tiny bubbles do not have time to find a new pressure equilibrium.
Besides global diffusion, strong conduction can thus promote minor stirring motions on small scales, preventing an abrupt decay of $A(k)_\delta$.
In this run, the spectrum slope in the inertial regime is steep, $A(k)_\delta\propto k^{-2/3}$, significantly different from the no-conduction run.
Radial profiles and SB$_{\rm x}$ maps (Fig.~\ref{fig:SBx}, bottom) are very similar to the $f=1$ model, retaining their initial spherical morphology.
The \Pt number is roughly 10 at the injection scale. Albeit turbulent regeneration starts to be more effective on large scales,
the key \Pt threshold appears to be an order of magnitude higher, as shown by the next model.

We suppress further the conductive flux by $f=10^{-2}$, a value advocated by several plasma physics theories (e.g.~\citealt{Rechester:1978, Chandran:1998}). The Prandtl number is 100 at the injection scale: turbulence can now restore part of the perturbations, though only near $L$ (the normalization rises again to $\sim$6 percent). This marked discrepancy between large and small scales induces a remarkably steep slope, $A(k)_\delta\propto k^{-4/5}$ ($E(k)\propto k^{-2.5}$), which should emerge in observed data in a clear way, if $f\sim10^{-2}$ is the conductive regime of the ICM. 
The $\delta \rho/\rho$ map (Fig.~\ref{fig:delta}, third panel) visualizes well the regeneration of turbulent eddies on large scales, while the small-scale flow remains considerably smooth, as corroborated by the SB$_{\rm x}$ map (Fig.~\ref{fig:SBx}, third row). 
Since this model shows a clear cutoff, it represents the cleanest case to retrieve the key threshold for the suppression of density perturbations, which we find to be \Pt$\,\sim100$. This is not a strict demarcation line, but rather a transition layer.

Only when conduction is substantially suppressed, $f=10^{-3}$ (the typically lowest suppression factor adopted in theories), the turbulent cascade is significantly restored, generating
the same peak and density spectrum down to $\sim$$L/2$.
Since thermal diffusion is too week, Kelvin-Helmholtz rolls and Rayleigh-Taylor instabilities can develop again over a large range, defining the entire flow dynamics (Fig.~\ref{fig:delta}) and perturbing the X-ray surface brightness (Fig.~\ref{fig:SBx}, second row).
Turbulent diffusion is able to efficiently mix the entropy profile, again lowering/increasing the central density/temperature (the discrepancy between $T_{\rm e}$ and $T_{\rm i}$ is now $\lta1$ percent; \S\ref{s:time}).
Conduction can affect only the scales smaller than 100 kpc, creating a gentle exponential decrease 
in the logarithmic $A(k)_\delta$. The suppression of $\delta$ reaches a factor of 2 near 30 kpc. Notice how conduction still dominates the diffusivity, overcoming (numerical) viscosity. 
When turbulent regeneration is efficient, it is not trivial to define an exact cutoff. Nevertheless, the threshold \Pt$\,\sim100$ ($l\sim100$ kpc) appears a robust criterium: at that scale we see the beginning of a substantial decay of the density spectrum (changing slope to $k^{-1/2}$).

\subsection{Mild turbulence: $M\sim0.5$} \label{r:med}

We now increase the level of turbulent motions by a factor of two, $M\sim0.5$ ($\sigma_v\sim750$ km s$^{-1}$). Turbulent energy is thus $\sim$14 percent of the thermal energy, still within the range retrieved by ICM observations and cosmological simulations. The characteristic eddy turnover time is $t_{\rm eddy}\sim0.8$ Gyr.

\begin{figure} 
    \begin{center}
       \subfigure{\includegraphics[scale=0.42]{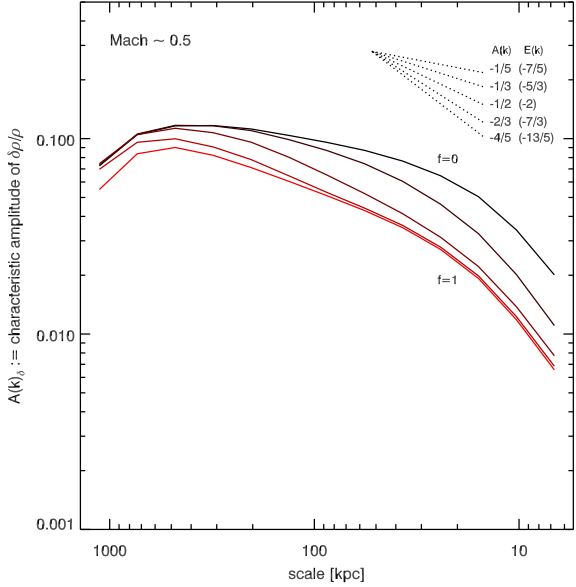}}
       \caption{Characteristic amplitude of $\delta \rho/\rho$, $A(k)_\delta=\sqrt{P(k)_\delta\,4\pi k^3}$, for the models 
       with mild turbulence $M\sim0.5$ and varying conduction. From top to bottom curve (black to bright red):
       $f=0,\, 10^{-3},\, 10^{-2},\, 10^{-1},\, 1$. The level of density perturbations is linearly related
       to the Mach number, $\delta \sim A(k)_{\delta, \rm max} \simeq 1/4\, M$. Strong conduction globally damps perturbations by a factor of 2$\,$-$\,$3, steepening the spectrum towards $\propto k^{-1/2}$ (from slightly shallower than Kolmogorov in the $f$$\,=\,$0 run). In the weakly conductive regime, the threshold for $A(k)_\delta$ decay is again $P_{\rm t}\sim 100$. 
      }
      \label{fig:Ak2}
     \end{center}
\end{figure}

Figure \ref{fig:Ak2} shows that the overall behaviour of $A(k)_\delta$ is similar to the previous set of models, with differences laying in the details. The purely turbulent case ($f=0$) forms the usual injection peak, with maximum at $\sim$12  percent (Table \ref{table:1}), i.e.~two times that of the previous run with half the turbulent velocity. Therefore, we infer that $\delta \rho/\rho \propto M$, or more precisely $\delta \rho/\rho \simeq 1/4\ M$ (with $l=L$; the 0.25 factor is likely related to $\gamma$ -- see \S\ref{s:time}). This is a key result that will be further corroborated by the runs with stronger turbulence and smaller injection scale. In the non-conductive runs, the direct relation between the density and velocity power spectrum holds also on intermediate scales.

The inertial range of $A(k)_\delta$ in the purely turbulent run is slightly shallower than the Kolmogorov spectrum ($\propto k^{-1/5}$; cf. \citealt{Kim:2005}). 
The decrease due to viscous dissipation occurs again within 15-20 kpc, while it is dominated by thermal diffusion in all the conductive runs. 
Given the significant level of turbulent diffusion, the entropy profile flattens twice more rapidly, doubling its central value to 700 keV cm$^2$ in less than 2 Gyr (the best-fit profile has almost twice lower normalization, with 20 percent larger core radius). Turbulent dissipational heating still plays a secondary role, since the temperature at large radii is increased by just 15 percent (its timescale is $\sim\,$$4\, t_{\rm eddy}$; \citealt{Gaspari:2013}). Overall, it is remarkable that the density perturbations follow in a fairly strict way the Kolmogorov velocities, even in the presence of a stratified atmosphere. 

We proceed testing thermal conduction, starting from the highest suppression factor.
As before, when $f=10^{-3}$, a gentle exponential cutoff develops. The Prandtl number is $\sim$2000 at the injection scale
(twice the previous models, \Pt$\propto M$),
signalling that turbulent regeneration is extremely efficient, as highlighted by the filamentary SB$_x$ image (Fig.~\ref{fig:SBx}, second column). The dynamics is dominated by K-H and R-T instabilities.  
We see a significant decline in density perturbations only at $l\lta60$ kpc, i.e.~\Pt$\sim 100$. This threshold appears again a robust indicator for the suppression of density inhomogeneities. 
When \Pt$<100$, the slope steepens from nearly Kolmogorov to Burgers spectrum ($A(k)_\delta\propto k^{-1/2}$).
The threshold and spectrum slope are independent of $L$ (see \S\ref{r:weak_half}), since shaped by the specific physical parameters of the plasma conductivity.

The effects of conduction can be best appreciated when
$f = 10^{-2}$. Using the key threshold \Pt$\sim100$, we predict a sharp decline around $\sim$330 kpc, and indeed we see the $k^{-1/2}$ spectrum appearing on this large scale. The suppression of $\delta$ on small scales is a factor of 3,
while near the injection scale turbulent regeneration is unhindered ($A(k)_\delta\sim11$ percent). The SB$_{\rm x}$ maps
also reveal that only the large filaments, edges, and rolls are retained. Assuming a deep exposure, X-ray imaging will thus help to assess the role of conduction and turbulence in the ICM.

Restoring the conductive flux to strong levels ($f \sim 0.1-1$, i.e.~\Pt$\sim20-2$ at $L$) promotes a global suppression of $\delta$, by a factor of $\sim$2$\,$-$\,$3, from large to small scales (Fig.~\ref{fig:Ak2}, red lines; the maxima of $A(k)_\delta$ show less separation compared with the models with weak turbulence). The slope is slightly shallower than Burgers spectrum, $A(k) \propto k^{-4/9}$ (although turbulence has now twice more strength). As in \S\ref{r:weak}, strong conduction can drive small-scale stirring, due to the fast conduction timescale relative to the gas/ion sound speed (Figure \ref{fig:time1}).
These models confirm that when \Pt$\lta20$ ($f\gta0.1$) the density perturbations are strongly suppressed, producing a similar $A(k)_\delta$ and smooth/spherical SB$_{\rm x}$ maps (Fig.~\ref{fig:SBx}), regardless of the exact suppression factor. 

A relevant result, which will be dissected in future work, is that conduction does not dramatically alter the power spectrum of the velocity field (nearly Kolmogorov), both in slope and normalization. Therefore, the quoted $M$ or velocity dispersion (roughly $v(k)$ at the injection scale) does not change between simulations with different suppression $f$.

\subsection{Strong turbulence: $M\sim0.75$} \label{r:strong}

In the next experiments, we test the case with strong (still subsonic) turbulence, with three times higher turbulent velocities compared with the reference model (\S\ref{r:weak}), $M\sim0.75$ ($\sigma_v \sim 1100$ km s$^{-1}$). Statistical steady state is reached fast, since $t_{\rm eddy}\sim0.5$ Gyr. The turbulent energy is $\sim0.31$ the thermal energy. Approaching the transonic regime becomes progressively unrealistic: it has been shown both in observations and cosmological simulations that the turbulent pressure support in the ICM should remain in the range few$\,$-$\,$30 percent (e.g.~\citealt{Churazov:2008, Lau:2009, Vazza:2009, Vazza:2011}), the upper envelope defined by unrelaxed systems. Dissipational heating also starts to be relevant ($\propto M^2$) increasing the overall temperature by $\gta\,$30 percent, though we are now interested in the relative variations. 

The current set of simulations corroborates the previous key findings.
According to our modelling, $\delta \rho/\rho\simeq1/4\,M$, we predict a characteristic level of density perturbations of 18.75 percent (for the weakly conductive models, $f<0.1$).
Measuring the normalization of the $A(k)_\delta$ peak ($f<0.1$), we retrieve a value of 18.8 percent (Table \ref{table:1}), an excellent match.
Albeit the evolution is strongly chaotic and nonlinear, it is remarkable that we can convert the Mach number and density perturbations in a linear and simple way.
In fact, the normalization of $A(k)_\delta$ 
is 2.9 (1.6) times that of the models with weak (mild) turbulence. 
Therefore, we expect real clusters to show density perturbations in the same range of the observed (low) Mach numbers.
This could also explain why clusters do not show exaggerated clumpiness. 

\begin{figure} 
    \begin{center}
       \subfigure{\includegraphics[scale=0.42]{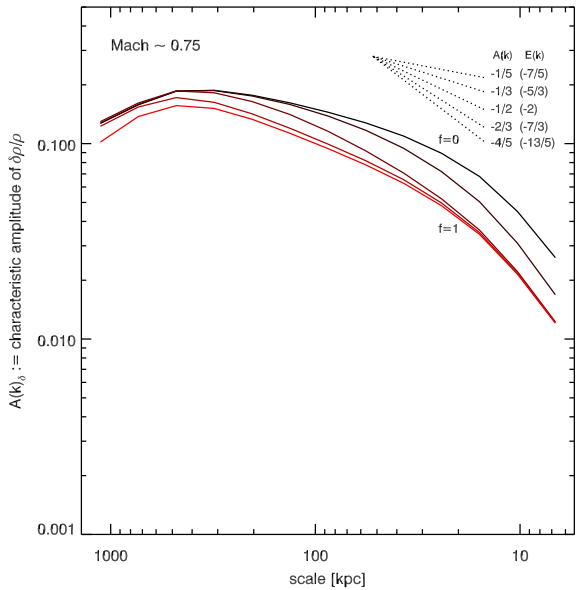}}
       \caption{Characteristic amplitude of $\delta \rho/\rho$, $A(k)_\delta=\sqrt{P(k)_\delta\,4\pi k^3}$, for the models 
       with strong turbulence $M\sim0.75$ and varying conduction:
       $f=0,\, 10^{-3},\, 10^{-2},\, 10^{-1},\, 1$. These models corroborate the previous findings: 
       the level of density perturbations is given by $A(k)_{\delta, \rm max} \simeq 1/4\, M$ (with $l=L$); 
       conduction damps density fluctuations by at least a factor of 2, steepening the spectrum towards $k^{-1/2}$
       (again from the Kolmogorov spectrum); 
       the threshold for the $A(k)_\delta$ decay is $P_{\rm t}\sim 100$.
       } 
     \label{fig:Ak3}
     \end{center}
\end{figure}  

The other features of the density power spectrum, i.e.~the slope and decay (Fig.~\ref{fig:Ak3}), follow the analysis described in the previous two sections. We summarize the key points.
The $A(k)_\delta$ slope of the $f=0$ run is again slightly shallower than $k^{-1/3}$, although not shallower than the $M\sim0.5$ model, indicating that the Kolmogorov spectrum is indeed the `saturated' regime, at least for subsonic turbulence\footnote{In the supersonic regime (\citealt{Federrath:2013}), the development of filamentary structures via strong shocks may further flatten the $\delta$ power spectrum (e.g.~\citealt{Kim:2005}).}.
Enabling conduction leads to the steeping of the characteristic amplitude towards the $k^{-1/2}$ regime, 
with the clearest manifestation in the $f=10^{-2}$ case,
where turbulence is able to regenerate perturbations near $L$ (notice the short formation of the gentle exponential decline), while small scales are substantially damped. 
In the strong conductive regime ($f\gta0.1$) conduction inhibits the perturbations up to a factor of $\sim$2. There is no drastic decline at small scales ($A(k)_\delta\propto k^{-4/9}$) since conduction also promotes minor stirring, albeit the increased gas sound speed/entropy (due to turbulent heating and diffusion) alleviates this process.
The commonly adopted suppression factors $f\ge0.1$ do not have a diverse impact on the density perturbations, which are damped over the whole range.

The turbulent Prandtl number is increased by a factor of 3 compared with the reference run, \Pt$\sim3-3000$ ($f\sim1-10^{-3}$) at the injection scale. This translates in a less marked -- though not globally different -- suppression between the purely turbulent and conductive case. 
More important,
\Pt$\sim 100$ corresponds to
$\sim$45 and 240 kpc for the $f=10^{-3}\ {\rm and}\ 10^{-2}$ model, respectively (in the $f\gta0.1$ runs, the transition is well over $L$). Indeed, the substantial decline occurs around these scales. This threshold is thus a robust criterium to estimate the impact of conduction in a realistic cluster atmosphere.

The surface brightness maps (Fig.~\ref{fig:SBx}; third column) highlight the substantial level of perturbations, which are now clearly manifest even in the cluster core (especially when $f\gta10^{-2}$). Such a configuration could be easily spotted in X-ray images even by eye inspection. Notice how in the weakly suppressed runs, the K-H and R-T instabilities translate into extended filaments and curved edges/fronts in X-ray brightness. The strong conductive flux instead smears out most of these features, preventing the ICM to depart from its spherical and smooth hydrostatic shape. 
This might be seen as an effective equation of state with a lower adiabatic index, implying a less reactive pressure during thermodynamic transformations ($P\propto \rho$ versus $P\propto \rho^{5/3}$).
Remarkably, the gas velocities remain $\sim$1000 km s$^{-1}$ on large scales, with the $v(k)$ spectrum not much affected by conduction (an in-depth comparison will be presented in a companion work).

\subsection{Small injection scale: $M\sim 0.25$, $L/2$} \label{r:weak_half}

The final set of simulations includes testing turbulence and conduction diminishing the injection scale. To carry on a proper comparison, we set the new injection scale as $L'=L/2\sim300$ kpc, but maintaining the turbulent
Mach number of the reference run, $M\sim0.25$ (via the scaling $\sigma_v\sim(N\,L'\,\epsilon^{\ast})^{1/3}$; \S\ref{s:turb}).
The characteristic eddy turnover time is $t_{\rm eddy} \sim 0.8$ Gyr, i.e.~the same characteristic timescale of the previous models with $M\sim0.5$ but larger injection scale (\S\ref{r:med}).
The typical turbulent energy is 3.5 percent of $E_{\rm th}$ (as in \S\ref{r:weak}).

\begin{figure} 
    \begin{center}
       \subfigure{\includegraphics[scale=0.42]{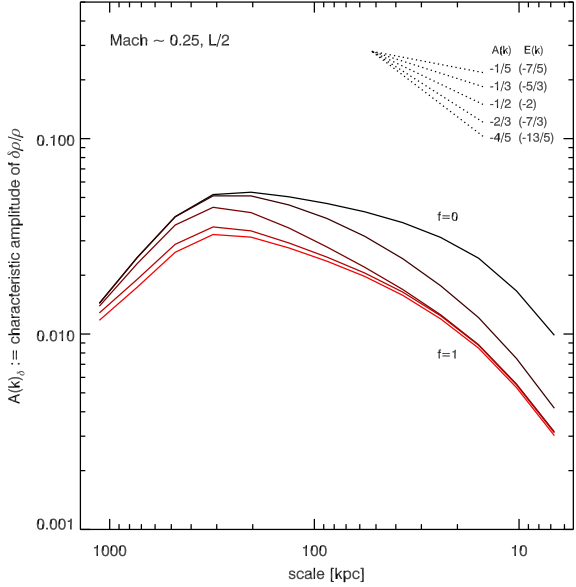}}
\caption{Characteristic amplitude of $\delta \rho/\rho$, $A(k)_\delta=\sqrt{P(k)_\delta\,4\pi k^3}$, for the models with weak turbulence $M\sim0.25$ and half the reference injection scale, $L'=L/2$. From black to bright red line, the suppression of conduction is $f=0,\, 10^{-3},\, 10^{-2},\, 10^{-1},\, 1$. The previous key features (e.g.~slopes and \Pt$\sim 100$ threshold) are valid also for this set of models. The system is dynamically similar to the runs with $M\sim0.5$, since the \Pt number (at injection scale) is the same ($t_{\rm turb}\sim0.8$ Gyr). Notice, however, that the $A(k)_\delta$ cascade is truncated at the new $L'$ and that the normalization is defined only by the Mach number (not $P_{\rm t}$).}       
        \label{fig:Ak4}
     \end{center}
\end{figure}

The key result is that the overall $A(k)_\delta$ spectra are analogous to the models with $M\sim0.5$ (Figure \ref{fig:Ak4}).
The characteristic Prandtl numbers, at the respective injection scales, are in fact the same. 
The major differences are twofold. First, the current spectra are truncated exactly at the smaller injection scale $L'$. 
Second, the normalization is defined {\it only} by the characteristic Mach number (and not $P_{\rm t}$).
In fact, using our previous $A(k)_\delta$ modelling and rescaling it with the new injection scale, we estimate $A(k)_{\rm \delta,max} \simeq 1/4 \,M \,(L'/L)^{1/5}\simeq0.054$. The measured peak in the simulated run is 5.3 percent (Table \ref{table:1}), confirming well the prediction. Notice that the $A(k)_{\rm \delta,max}$ fit is very weakly dependent on the injection scale, implying that the linear conversion between $\delta$ and $M$ is fairly general (see \S\ref{s:time}). 

Similar to previous $M\sim0.5$ models, the strongly conductive runs ($f\gta0.1$) show density perturbations damped by a factor of 2$\,$-$\,$3, from large to small scales. The $A(k)_\delta$ slope follows also the previous trends, slightly shallower than Kolmogorov in the $f=0$ run ($\propto k^{-1/5}$; higher $D_{\rm turb}$ means faster $\delta$ regeneration), while steepening towards the Burgers-like spectrum for the conductive runs.
Using the threshold \Pt$\sim100$, we retrieve the region where $A(k)_\delta$ substantially decays. Interestingly, 
the estimated cutoff for the $f=10^{-2}$ run is slightly larger than the injection scale ($\sim$330 kpc) and indeed we see
the decrease in normalization (by 10 percent), instead of a progressive decline from the $f=0$ baseline.
In the $f=10^{-3}$ run, the threshold is still below 100 kpc, allowing the exponential cutoff to partially develop, although 
entering quickly the regime of substantial suppression on few tens kpc. 
The hydrodynamics is similar to the $M\sim0.5$ models, with K-T and R-T instabilities producing filaments and edges in SB$_x$ (if $f<0.1$), but now with maximum size $\sim$100 kpc. On the other hand, strong conduction has the ability to preserve the smooth and spherical shape of the galaxy cluster (cf. last row in Fig.~\ref{fig:SBx}).

\section{Discussion}  \label{s:disc} 
We now explain and discuss the key features of the models,
and then constrain ICM conduction and turbulence in real Coma cluster through density perturbations.
 
\subsection{Dominant timescales \& simple predictions}\label{s:time}
\begin{figure} 
    \begin{center}
       \subfigure{\includegraphics[scale=0.35]{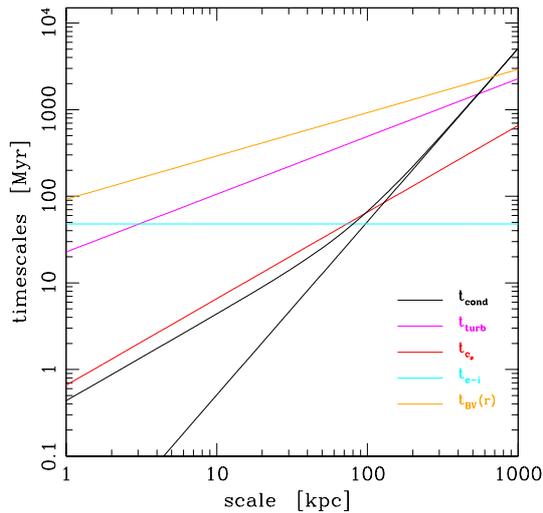}}
       \caption{Typical timescales in the core of Coma ($n_{\rm e}$ and $T_{\rm e}$ are roughly constant) as function of $l$, for the run with $M\sim0.25$ and $f=1$: conduction (black; Eq. \ref{e:tcond}), turbulence (magenta; Eq. \ref{e:tturb}), sound (red; $t_{c_{\rm s}}=l/c_{\rm s}$), electron-ion equilibration (cyan; Eq. \ref{e:texc}).
       The conduction timescale for the saturated zones corresponds to the upper black envelope, limited by the electron sound speed ($\simeq$43 times that of protons). The yellow line is the Brunt-V\"ais\"al\"a (buoyancy) timescale, 
       as function of $r$, using an average 
       $g\sim5\times10^{-9}$ cm s$^{-2}$ and shallow core entropy slope ($\sim0.1$); $t_{\rm BV}$ is similar to the free-fall time, within a factor of a few.
      \label{fig:time1}}
     \end{center}
\end{figure}

The qualitative evolution of a dynamical system can be described via the timescale ratio of the physical processes involved,
i.e.~via characteristic dimensionless numbers. Let us consider as reference Figure \ref{fig:time1}, showing the typical timescales for $M\sim0.25$ and strong conduction ($f=1$) run, within the core radius. 
The first key quantity is the (turbulent) Mach number,
$M=t_{\rm c_s}/t_{\rm turb}$ ($<1$), which we found to be linearly related to the relative density perturbations $\delta$. The scaling may be justified by simple physical arguments (see also Zhuravleva
et al.~2013, in prep.). Perturbations tend to settle back towards shells of similar entropy.
Assuming an isobaric blob, its entropy varies as $K\propto \rho^{-\gamma}\propto (1+\delta)^{-\gamma}$ -- for an isothermal blob $K\propto(1+\delta)^{1-\gamma}$. 
Since the cluster entropy typically scales with the radius, the displacement is directly linked to the density contrast, $\Delta r /r_{\rm init} \approx \Delta K/K_{\rm init} = 1-(1+\delta)^{-\gamma}\approx \gamma\, \delta$. In a stratified medium,
the balancing force is $F\propto g\, \delta$, again directly related to the displacement. Since the specific kinetic energy is $v^2 \propto F\,\Delta r\propto (\Delta r)^2$, we infer that $M \propto \Delta r \propto \delta$, with the normalization shaped by the effective $\gamma$ and the $K(r)$ slope.
In the limiting case of negligible buoyancy,
density may be instead roughly treated as a passive scalar.
In the classic theory of Obukhov \& Corrsin (\citealt{Warhaft:2000} for a review),
the spectrum of passive scalars is expected to be directly related to $v(k)$.
The scaling $\delta \propto M$ is thus fairly general.

The second key quantity is the conduction timescale (Fig.~\ref{fig:time1}),
which is typically the shortest one ($\lta 50$ Myr). 
The black line shows that conduction dominates the dynamics over the whole range of scales,
even in the saturated regime (upper black line). 
Increasing the level of suppression ($f \ll 1$) induces the crossover of the conduction and turbulence (magenta) timescale. 
The predicted threshold for the turbulent regeneration of perturbations is $P_{\rm t}\equiv t_{\rm cond} / t_{\rm turb}\sim 100$. If $P_{\rm t}\sim100$ occurs within the injection scale, the normalization of the spectrum does not decrease, and a decay appears on intermediate scales (Table \ref{table:1}). 
In this work, we used a turbulent diffusion coefficient $c_{\rm t}=1$, yet the actual value could be as low as 0.1$\,$-$\,$0.01 (\citealt{Kim:2003, Dennis:2005}), rescaling the threshold to $P_{\rm t}\sim 1$.

Other timescales, linked to the relevant physics, are involved in the complex dynamics, although inducing secondary effects.
The Brunt-V\"ais\"al\"a timescale tracks the impact of the restoring buoyant forces along the radial direction:
$t_{\rm BV}=[\gamma \,(r/g)\, ({\rm d}\,{\ln K}/ {\rm d}\,{\ln r})^{-1}]^{1/2}$. 
Buoyancy is thus defined by gravity $g$ and the slope of the entropy profile\footnote{In the case of anisotropic conduction, buoyancy is constrained by ${\bf \nabla} T$ (\citealt{Sharma:2009_proc}), which is usually shallower than ${\bf \nabla} K$.}.
Within the core radius, ${\bf \nabla} K$ is very shallow (and gravity roughly constant), hence $t_{\rm BV} > t_{\rm turb}$, or Froude number $Fr > 1$ (Fig.~\ref{fig:time1}, yellow line), implying that the ICM is approaching neutral buoyancy and turbulent motions can easily retain isotropy. On large scales, the entropy profile follows the self-similar slope $\propto r$, as most clusters, but it is balanced by the 
decreasing gravity: buoyancy can again not efficiently inhibit radial motions. Substantially weaker stirring
may progressively force the chaotic flow to follow tangential streamlines ($Fr \ll 1$).
Notice that the buoyancy timescale is essentially the free-fall time, within a factor of a few: $t_{\rm ff} = (2 r/g)^{1/2}$. The ratio of the free-fall time and the gas sound-crossing time (red line) is always $> 1$, i.e.~the cluster is in global hydrostatic equilibrium. As seen in Fig.~\ref{fig:time1}, in the strongly conductive runs ($f\gta0.1$) the sound-crossing time may slightly fall below the conduction time, approaching the small scales. Albeit saturation prevents electrons to conduct faster than their thermal speed, $t_{c_{\rm s}}/t_{\rm cond}\lta 1$ warns that small fluctuations do not have time to re-adjust via pressure equilibrium. Progressively stronger conduction is thus not able to completely stifle $A_\delta$ at large $k$. 

The electron-ion equilibration introduces another relevant timescale (Fig.~\ref{fig:time1}, cyan). On small scales, $t_{\rm e-i}$ is usually larger than $t_{\rm cond}$ and comparable to the eddy turnover time. This means that the proton temperature can not become as quickly isothermal as $T_{\rm e}$ (lagging by 50$\,$-$\,$100 Myr), a feature aggravated by the
continuously chaotic ambient. 
The discrepancy resides in the range 1$\,$-$\,$15 percent, from the inner to outer radial bins (for model $M\sim0.25$, $f=1$),
decreasing to a maximum of 10, 3, and 1 percent, for $f=10^{-1}, 10^{-2}$, and $10^{-3}$, respectively. Higher Mach number linearly increases the maximum discrepancy (occurring at $r > r_{\rm c}$, where the medium is more rarefied). 
The estimate of $|T_{\rm e}-T_{\rm i}|/T_{\rm i}$ can be used to properly constrain the turbulent velocities from the observed spectral lines (e.g.~\citealt{Inogamov:2003}). In order to remove thermal line broadening ($\propto T_{\rm i}^{1/2}$), it is assumed $T_{\rm i} = T_{\rm e}$, hence an effectively lower ion temperature would imply higher turbulent $\sigma_v$. 

\subsection{Modes of perturbations: isobaric, isothermal, isentropic}\label{s:modes}
It would be tempting to characterize chaotic fluctuations with a single mode, as adiabatic (isentropic), isobaric, or isothermal.
As experiment, we set up the cube with pure isobaric fluctuations (fixing Kolmogorov spectrum). 
The hydro run develops the turbulent cascade, preserving the isobaric fluctuations, which gradually decay in few $t_{\rm eddy}$ due to no forcing.
However, as soon as conduction is enabled, 
the mode of perturbations suddenly change from isobaric to isothermal/adiabatic. The result is the generation of acoustic oscillations and, strikingly, a density power spectrum similar to that of CMB (\citealt{Planck:2013_spectrum}), although the two processes are inherently different -- CMB evolution is driven by (compressive) gravity. 
The spectra observed in clusters 
do not present such an imprint, meaning that the actual type of fluctuations is a {\it mixture} of desynchronized modes between the adiabatic and isobaric extremes\footnote{Cosmological runs also show mixed perturbations (\citealt{Zhuravleva:2013}), further corroborating the realism of our turbulence driving.}. 

We analyze our models 
to understand the correlation between relative density ($\delta_\rho$) and temperature fluctuations ($\delta_T = T_{\rm e}/T_{\rm b} -1$, where $T_{\rm b}$ is  the underlying radial profile).
The correlation is positive for adiabatic fluctuations ($\delta_\rho/\delta_T =\gamma-1$; constant entropy),
while negative for the isobaric mode ($\delta_\rho/\delta_T = -1$); the isothermal mode is the intermediate regime ($\delta_\rho/\delta_T = \infty$). Analyzing the linear regression coefficient of the $\delta_T$-$\delta_\rho$ diagram ($512^3$ points) is not much meaningful, since there is too high dispersion. 
We use thus Pearson coefficient\footnote{The correlation coefficient is the ratio of the covariance of two populations to the product of their standard deviations; $r_{\rm P}\in [-1,\,1]$.} to assess the degree of positive and negative linear correlation. In the no-conduction run with $M\sim0.25$, the correlation is $r_{\rm P}\simeq-0.86$, implying that weak turbulence prefers isobaric fluctuations (adiabatic sound waves are negligible). Since stirring is subsonic, the fluctuations can not deviate much from hydrostatic equilibrium.
When conduction is enabled, though partially suppressed, the scatter in the phase diagram progressively increases: fluctuations start filling also the adiabatic region, showing $r_{\rm p}=-0.82, -0.78, -0.65$, for the runs with $f=10^{-3},10^{-2},10^{-1}$, respectively. With full Spitzer conduction ($f=1$), the correlation drops to $r_{\rm p}=-0.30$: the perturbations are now almost equally isobaric and adiabatic. 

Increasing the strength of stirring ($M\sim0.75$), rises the level of turbulent mixing, leading to shallower entropy gradients.
In fact, even in the hydro run, the adiabatic and isobaric modes are balanced
($r_{\rm P}\simeq-0.20$). Conduction tries to push the correlation towards the isothermal regime, albeit there is no clear preference in the mixture of perturbations
($r_{\rm P}\sim-0.1$; $f=1$ run).
Smaller injection scales also induce a more efficient turbulent cascade, promoting mixing and leading to higher adiabatic fluctuations (the latter are also enhanced in supersonic\footnote{In the presence of strong shocks, the correlation can overcome the adiabatic limit, $\delta_\rho/\delta_T\ll\gamma-1$.} turbulence).
Overall, we suggest to exploit the correlation between $\delta_T$ and $\delta_\rho$ 
to assess the dominant physics in the ICM.

\subsection{Comparison with real Coma cluster}\label{s:comp}
In order to constrain the transport properties in the bulk of the ICM,
we take as exemplary case Coma cluster, due to the availability of new deep observations ($\sim$500 ks; Churazov et al.~2013, in prep.). 
We refer to \citet{Churazov:2012} for the extraction method of $A_\delta$, accounting for Poisson noise, point sources, and deprojection. The hot ICM, as in Coma, is best suited for this analysis, due to the potential strong conductivity
and negligible cooling or AGN heating. 
In cold, dense systems ($T_{\rm vir}< 5$ keV) conduction is at least 100$\,$-$\,$1000 times weaker, thus its impact is expected to be marginal (\citealt{Dolag:2004,Smith:2013}), even if unsuppressed, and very difficult to isolate (the analysis of other systems will be provided in a subsequent work).

The previous simulations provided a clear picture of the interplay between conduction and turbulence in the hot ICM. We can thus formulate a simple, general model linking conduction and turbulence to the
the level of density fluctuations at a given physical scale $l$ ($=k^{-1}$):
\begin{equation}\label{e:fit}
\delta(l) \simeq\left[0.25 \,M \,\left(L_{\rm inj}/{L_{500}}\right)^{\alpha_{\rm h}}\right]\; (l/L_{\rm inj})^{\alpha_{\rm c}}.
\end{equation}
The term in brackets represents the normalization of $\delta$ which must be rescaled to the current injection scale $L_{\rm inj}$, using the slope given by the purely hydrodynamic cascade, $\alpha_{\rm h}$ ($\approx$ 0.2$\,$-$\,$0.3;
Table \ref{table:1}). As shown before, the characteristic spectrum slope $\alpha_{\rm c}$ is dictated by the level of conduction, and can be much steeper than $\alpha_{\rm h}$ ($\alpha_c\gta0.45$, for $f\gta10^{-2}$). 
A steep slope combined with globally smooth
residual/SB$_{\rm {x}}$ maps indicates that 
the normalization should be reduced by 1.3$\,$-$\,$2, due to very strong conductivity ($f\gta0.1$; Table \ref{table:1}). 
This prescription is sufficient to assess the physical state of the hot ICM.
For a more precise modelling, 
we suggest to insert 
two (exponential) cutoffs linked to the injection, $\exp\left[-\left(l/a_1 L\right)^{\eta_1}\right]$, and dissipation scale, $\exp\left[-\left(a_2 \Delta x/l \right)^{\eta_2}\right]$;
typical parameters fitting our simulated spectra are $\eta_1=4$, $\eta_2=3/2$, $a_1=2$, $a_2=3$.

\begin{figure} 
    \begin{center}
       \subfigure{\includegraphics[scale=0.42]{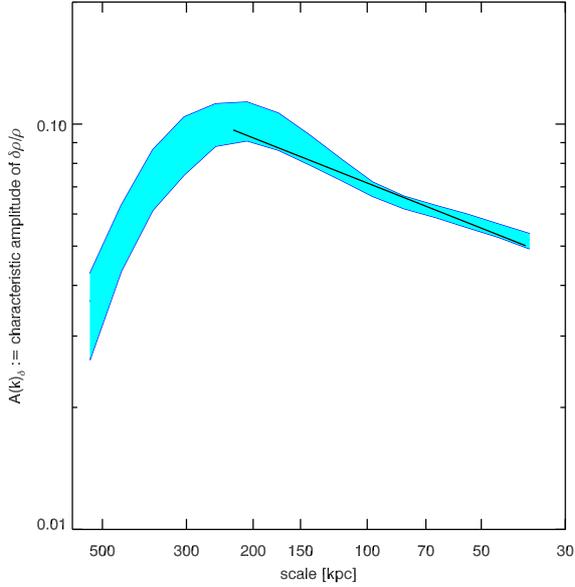}}
       \caption{Cyan envelope: characteristic amplitude of $\delta \rho/\rho$ in real Coma, extracted from deep {\it Chandra} observations (Churazov et al.~2013, in prep.; $\sim$500 ks), within the statistical uncertainties. 
       Black line: $A_\delta(k)$ prediction based on our modelling. The state of the ICM is best matched by our simulations with highly suppressed conduction, $f\simeq10^{-3}$, since the spectrum shows a shallow slope, $\alpha_{\rm c}\simeq0.36$, slightly steeper than Kolmogorov. The normalization (10 percent) indicates that the level of turbulence is $M\simeq0.45$. The spectrum also suggests an injection scale of $\sim$250 kpc, although the precise value is uncertain due
       to the selection of the unperturbed model.
       } 
     \label{fig:comp}
     \end{center}
\end{figure}

In Figure \ref{fig:comp}, we present the characteristic amplitude of density perturbations observed in real Coma cluster
(within statistical uncertainties; cyan envelope), compared with the prediction based on our modelling (black line).
It is clear that the observed spectrum is shallow. A power-law with slope $\alpha_{\rm c}\simeq0.36$, slightly steeper than Kolmogorov (0.33), fits well the inertial range of the spectrum.
Coma observations of pressure perturbations further corroborate this trend (\citealt{Schuecker:2004}). 
Any plasma with significant conduction would produce instead a steeper spectrum. We can therefore exclude a strong or mild conductive state of the ICM, $f\ge10^{-2}$. 
The inertial range develops unimpeded down to tens kpc, indicating that turbulent regeneration is substantial and that the Prandtl number at the injection scale is high ($P_{\rm t}\gta 2000$). The fact that we do not see a sharp cutoff, but only a gentle decay, 
indicates that the suppression factor is very low, $f\simeq10^{-3}$, although not $f=0$ (the non-conductive cascade would be significantly higher below 60 kpc; \Pt$\sim100$).
In addition, the observed $\delta$ map (\citealt{Churazov:2012}, 2013) is not smooth, revealing a variegated morphology of local features, reminiscent of Fig.~\ref{fig:delta} (top two panels). The relation between the velocity and density spectrum can finally break any ambiguity of the proper regime (Gaspari et al.~2014, in prep.).

As reviewed in \S\ref{s:intro} and \ref{s:MHD}, the level of suppression in turbulent plasmas is a matter of debate. It strongly depends on the topology of magnetic field lines and the relation between the electrons mean free path and the field correlation length, $l_B$ (e.g.~\citealt{Rechester:1978,Chandran:1998,Malyshkin:2001_mirror,Narayan:2001, Chandran:2004}). For example, the latter authors suggested a suppression $f\sim0.1$ if $\lambda_{\rm e} \ll l_B$, while in the opposite regime 
magnetic mirrors can further suppress heat transport. On top of this, kinetic plasma instabilities can substantially lower $f$.
If the field is more ordered, the conductivity across the field can be heavily suppressed, $f\ll1$, as found by observational and theoretical analyses of sharp features in the ICM, e.g.~cold fronts (\citealt{Ettori:2000,Roediger:2013,ZuHone:2013}) and linear filaments (\citealt{Forman:2007, Sanders:2013_fil}).
Interestingly, the low value of $f$ could explain the survival of positive $T$ gradients in the center of many clusters since $z\sim1.2$ (\citealt{McDonald:2013}). It is also clear that such a low conductivity can not provide sufficient heating for quenching radiative cooling, leaving AGN feedback as the main solution for the cooling flow problem (\S\ref{s:intro}).

The $A_\delta(k)$ peak, i.e.~the normalization, is $\simeq\,$10 percent,
which can be achieved by a level of turbulence with $M\simeq0.45$. The ratio of turbulent to thermal energy is thus 11 percent, concordant with observations and cosmological simulations of clusters (\S\ref{s:turb}).
The spectrum suggests an injection scale $\sim$250 kpc, similar to the estimated impact parameter of merging clusters
(\citealt{Sarazin:2002}). 

We warn that the observed low frequencies have significant uncertainty, due to the stochastic nature of perturbations and modelling errors, as the deprojection and the underlying profile removal. For instance, \citet{Churazov:2012} tested  asymmetric models, finding that non-sphericity can vary the deprojected amplitude above 300 kpc up to a factor of 2, while smaller scales remain unaffected.
The current simulations also do not model mergers or external accretion. 
Density inhomogeneities due to the substructures in
the potential and accreted filaments may come on top of the large-scale 
perturbations directly related to the turbulent cascade 
(\citealt{Churazov:2012,Sanders:2013_fil}). 
Multiple injection scales may also flatten the spectrum, although they require fine-tuning to be masked as a single spectral profile.
While the turbulent cascade can be formed by a variety
of drivers, its appearance, modelled in the present work, is nevertheless expected to
be largely universal. It is no coincidence that the cascade in cosmological simulations is nearly Kolmogorov (e.g.~\citealt{Vazza:2011}), similar to our hydro runs. We remark that strong conduction would be clear in the spectrum, regardless of the above-mentioned mechanisms, uniquely stifling the cascade.

On very small scales, the observed spectrum is limited by Poisson noise, overcoming the fluctuations below $\sim$30 kpc.
The range 30$\,$-$\,$300 kpc is yet sufficient to determine the state of the ICM, since the effects of diffusion would clearly emerge within this range (\S\ref{s:res}). A steep slope can indeed not be produced by changing $M$ or $L_{\rm inj}$, but only increasing $f$. The SB$_{\rm x}$ map can further clarify the proper regime.
Notice that if the precise $L_{\rm inj}$ is larger, the estimate of $M$ should not vary significantly, since the more extended cascade provides a higher $A_\delta(k)$ peak. With even deeper data, the $A_\delta(k)$ slope could 
be traced down to smaller scales,
further constraining the role of conduction (notice that low resolution smoothens the $\delta$ field). 
Future deeper X-ray observations might further improve these values, perhaps combining several clusters,
and providing a high-precision spectrum of ICM perturbations.

In passing, we note that the $\delta$ prescription (Eq.~\ref{e:fit}) can be very useful for those (subgrid) simulations, 
semi-analytic or analytic works, which intend to model, ab initio and in a simple way, the interplay of conduction, turbulence, and inhomogeneities in the ICM. For instance, the above equation can be used either to set the level of density (and $T$; \S\ref{s:modes}) perturbations from the Mach number (particularly useful for clumpiness studies), {\it or} to impose the effective Mach number from the amplitude of density fluctuations.
The current method can be readily applied to other areas of astrophysics, as the interstellar or intergalactic medium.
Interestingly, also the ISM shows a nearly Kolmogorov slope over five orders of magnitude (\citealt{Armstrong:1995}), though steepening in dwarf galaxies (\citealt{Dutta:2009}); conduction seems thus substantially 
suppressed in different astrophysical atmospheres.

\section{Conclusions}\label{s:conc}
Using 3D high-resolution hydrodynamic simulations, we thoroughly tested the physics of thermal conduction ($f=0\rightarrow1$) and turbulence ($M=0.25 \rightarrow 0.75$)
in the hot intracluster medium, tracking both electrons and ions (|$T_{\rm e}-T_{\rm i}|/T_{\rm i}\lta 0.15$). 
We showed how to exploit the power spectrum of the relative gas density perturbations $\delta=\delta\rho/\rho$
(normalization, slope, decay),
in order to accurately constrain the effective conductive and turbulent state of the ICM.
We chose Coma cluster as reference laboratory.
The main results are as follows.

\renewcommand{\labelitemi}{$\bullet$}
\begin{itemize}

\item
The {\it normalization} of the characteristic amplitude spectrum, $A(k)_\delta=\sqrt{P(k)_\delta\,4\pi k^3}=\sqrt{E(k)_\delta \, k}$, determines the strength of turbulence, and vice versa.
The peak amplitude is {\it linearly} related to the average 3D Mach number as $A(k)_{\delta, \rm max} = c\, M$, where $c \simeq 0.25$ 
for injection scale $L_{\rm inj} \simeq 500$ kpc.
In the hydro run, the spectrum of $\delta$ tracks that of velocities even on intermediate scales.
After rescaling, the steady spectra are remarkably similar despite different $M$ or $L_{\rm inj}$, allowing to build a general theory of ICM perturbations. \\

\item 
The {\it slope} of $A_\delta(k)$ determines the level of diffusion dominated by conduction. In a non-conductive ICM ($f=0$), subsonic stirring motions generate a nearly {\it Kolmogorov} cascade, $E_\delta(k)\propto k^{-5/3}$. Similar to velocities, the inertial range of density perturbations peaks at the injection scale, and decays below 10 kpc due to `viscous' dissipation.
Increasing the level of conduction, with magnetic suppression $f=10^{-3} \rightarrow 1$, progressively {\it steepens the  spectrum slope towards the Burgers-like regime} ($E_\delta(k)\propto k^{-2}$; Table \ref{table:1}), a feature that would be manifest in observations. The velocity spectrum is not much affected by conduction.
Even with $f=1$, the $\delta$ amplitude does not drop to zero,
since strong conduction also promotes fluctuations.
The $A_\delta(k)$ slope is only weakly dependent on $M$, becoming slightly shallower above $M\gta\,0.5$.\\

\item
The dominant dimensionless number or timescale ratio, shaping the flow dynamic similarity (\S\ref{s:time}), is the turbulent Prandtl number: \Pt$ = D_{\rm turb}/D_{\rm cond}=t_{\rm cond}/t_{\rm turb}$. The threshold $P_{\rm t}\lta100$  approximately indicates where $A_\delta(k)$ has a significant {\it decay} due to diffusion.
The transition is very gentle for strong suppression of conduction, $f\lta10^{-3}$, becoming a sharp decay -- though not a cutoff -- in the opposite regime. If $P_{\rm t}\sim100$ occurs above the injection scale, density perturbations are inhibited over the whole range of scales, inducing a decrease in normalization up to a factor of $\sim$2 (on small scales the suppression can reach a factor of 4). This state occurs only with strong conductivity, $f\ge0.1$ 
and would be pinpointed by the SB$_{\rm x}$ or residual $\delta$ images, in which K-H/R-T rolls and filaments are washed out, preserving the smooth and spherical shape of the cluster. \\

\item
Realistic perturbations, in a stratified system, are characterized by a mixture of modes, shaped by
the dominant physics. 
Weak turbulence drives higher isobaric perturbations ($\delta_\rho/\delta_T = -1$); strong turbulence enhances the adiabatic modes ($\delta_\rho/\delta_T = \gamma-1$), while increasing conduction forces both modes towards the intermediate isothermal regime. Although density statistics is much better constrained by observations, unveiling temperature perturbations could substantially advance our knowledge of the ICM physics. \\

\item
Based on our experiments, we provided a general analytic model to constrain density perturbations, conduction, and turbulence in the {\it bulk} of the ICM:
$\delta(l) \simeq[0.25 \,M \,(L_{\rm inj}/{L_{500}})^{\alpha_{\rm h}}]\; (l/L_{\rm inj})^{\alpha_{\rm c}}$ (see \S\ref{s:comp}).
We apply it to new very deep {\it Chandra} observations of Coma. The observed $A_\delta(k)$ spectrum is best consistent with strongly suppressed conduction, $f\simeq10^{-3}$, and mild subsonic turbulence, $M\simeq0.45$ ($L_{\rm inj}\sim250$ kpc). 
The latter ($E_{\rm turb}\simeq0.11\, E_{\rm th}$) is in agreement with cosmological simulations of clusters formation and line-broadening observations.
The low conductivity corroborates the survival of local sharp features (cold fronts, filaments, bubbles), and indicates that cooling flows may not be balanced by conduction, leaving AGN feedback as the main driver of heating.

\end{itemize}

The increasing quality and sample size of future X-ray data will provide a key opportunity to exploit this new spectral modelling, and hopefully to open the path to high-precision physics of the ICM, as has been done for the CMB.

\section*{Acknowledgments}
The FLASH code was in part developed by the DOE NNSA-ASC OASCR Flash center at the University of Chicago. 
We acknowledge the MPA, RZG, and CLS center for the availability of high-performance computing resources. 
MG is grateful for the financial support provided by the Max Planck Fellowship.
EC acknowledges support by the Research Program OFN-17 of the Division of Physics, Russian Academy of Science.
We thank R. Sunyaev, A. Schekochihin, F. Vazza, and the anonymous referee for helpful comments.

\bibliographystyle{aa}
\bibliography{biblio}

\appendix

\section{Mexican Hat versus FFT spectrum} \label{app:1}
Since the data cube is non periodic, computing the power spectrum via fast Fourier transforms is in principle 
inconsistent. 
As a consequence, the unresolved large scale power can leak into the available frequency range, distorting the spectrum. 
We use thus a modified $\Delta$-variance method,
known as `Mexican Hat' filtering (MH; cf.~\citealt{Arevalo:2012}). For each spatial scale $\sigma$, the method consists of three steps:
\begin{enumerate}
\item
the real-space cube $C$ is convolved with two Gaussian filters having slightly different smoothing lengths: $\sigma_1=\sigma/\sqrt{1+\epsilon}$
and $\sigma_2=\sigma \sqrt{1+\epsilon}$, where $\epsilon\ll1$; 
\item
the difference of the two cubes is computed, resulting in a cube dominated by the fluctuations at scales $\approx\,$$\sigma$ (the difference of two Gaussian filters is simply the Mexican Hat filter, $F(x)\propto \epsilon\,[1-x^2/\sigma^2]\,\exp[-x^2/2\sigma^2]$, characterized by a positive core and negative wings);
\item
the variance $V_\sigma$ of the previous cube is calculated and recast into the estimate of the power, knowing that
\begin{equation*}
V_\sigma = \int (C \ast F)^2\; {\rm d}^3 x =\int P_k \,\left|\hat{F}_k\right|^2\; {\rm d^3} k \propto P_\sigma \,k^3. 
\end{equation*}
\end{enumerate}
We refer to \citet{Arevalo:2012} -- appendix A -- for the technical procedure and normalization.
In order to handle the non-periodicity of the cube, we use a `mask' which is 1 inside and 0 outside the domain. 
The big advantage of the MH is that it avoids any leakage of power linked to the non-periodicity of the data;
the drawback is that it can not capture very sharp features in the power spectrum, due to the smoothening over $\Delta k\sim k$. 

\begin{figure} 
    \begin{center}
       \subfigure{\includegraphics[scale=0.38]{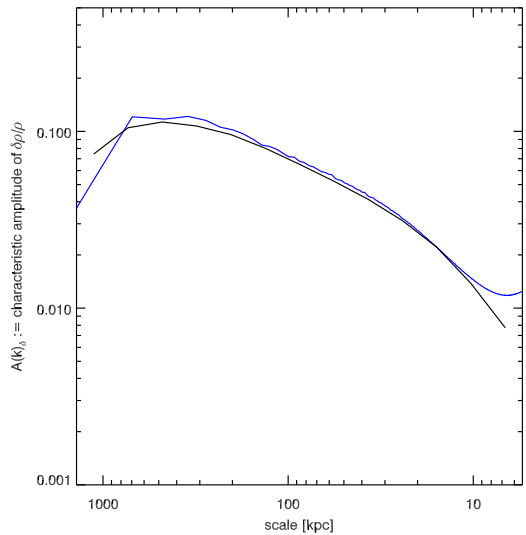}}
       \caption{Comparison of the characteristic amplitude spectra (for the run with $M\sim0.5$ and $f=10^{-2}$), computed with two different methods: Mexican Hat filtering (black) and fast Fourier transforms (blue). The retrieved spectrum is consistent in both cases, without major differences.} 
              \label{fig:Ak_comp}
     \end{center}
\end{figure}

In Figure \ref{fig:Ak_comp}, we show the comparison between the MH and FFT spectrum, for the run with $M\sim0.5$ and $f\sim10^{-2}$. In our study, there is no dramatic difference between the two methods.
The slope in the inertial range is almost identical. 
At very small scales,
the FFT spectrum produces a characteristic hook, in part due to the the numerical noise near the maximum resolution, but also due to the contamination of jumps at the non-periodic boundaries. The MH spectrum shows instead a gentle decline.
In the opposite regime, the MH filter tends to smooth the scales greater than the injection scale,
while the FFT spectrum shows a steeper decrease. The FFT peak is slightly higher, typically by 2$\,$-$\,$3 percent,
likely affected by the non-periodic box. Progressively trimming the box 
increases the relative normalization of the FFT spectrum, even by 20 percent, while distorting the low-frequency slope;
the MH spectrum is instead unaltered.

\section{$\beta$-profile in Fourier space} \label{app:2}
We present here the analytic conversion of the $\beta$-profile to Fourier space, 
and its interplay with a power-law noise.
Using the notation $\hat{f}(k)=\int^{\infty}_{-\infty}f(x)\,\exp[i k x]\, {\rm d} x/\sqrt{2\pi}$,
the Fourier transform of the $\beta$-profile (Eq.~\ref{nComa}) results to be
\begin{equation}\label{e:Pbeta}
\hat{n}_\beta(k) = n_0\, \frac{2^{1-\xi}\; r_{\rm c}^{\xi+1/2}\; |k|^{\xi-1/2}\; K_{1/2-\xi}[|k|\,r_{\rm c}]}{\Gamma[\xi]},
\end{equation}
where $\xi\equiv3\beta/2$, $K_{1/2-\xi}$ is a modified Bessel function of the second kind, and $\Gamma$ is the Gamma function.
The (1D) power spectrum is as usual retrieved as $P(\hat{n}_\beta)=|\hat{n}_\beta(k)|^2$.
Assuming $\beta=2/3\simeq0.66$ (a typical value for galaxy clusters), 
the power spectrum of the $\beta$-profile reduces to
\begin{equation}\label{e:Pbeta_simp}
P(\hat{n}_\beta)= \left(\frac{\pi}{2} r_{\rm c}^2 \right) \,\exp[-\,2\, |k|\, r_{\rm c}].
\end{equation}

\begin{figure} 
    \begin{center}
       \subfigure{\includegraphics[scale=0.55]{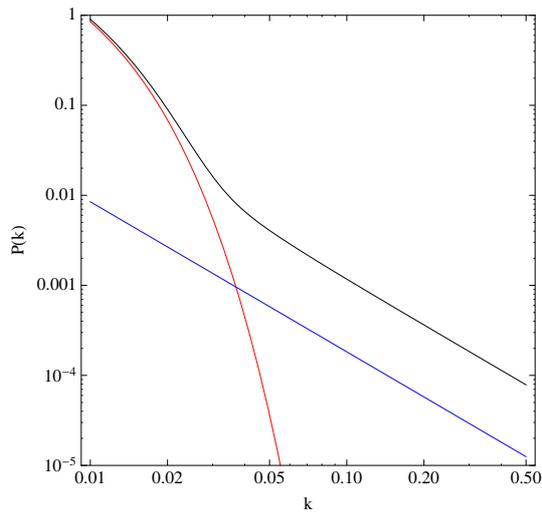}}
       \caption{Analytic 1D power spectra: $\beta$-profile (red), Kolmogorov noise (blue), $\beta$-profile perturbed by the noise (black; $P(\hat{n}_{\rm p})$). The spectrum is normalized to the value at $k_0=1/L=0.01$ (dimensionless units; $2\pi$ is dropped for clarity).
The core radius is $r_{\rm c}=20$, i.e.~$L/5$. 
The relative amplitude of the noise is $\sim$10 percent.  
The noise clearly emerges beyond the core radius ($k > 0.05$), regardless of 
large-scale structures.}
              \label{fig:beta}
     \end{center}
\end{figure}

The previous equation strikes for its simplicity, and can be readily used in semi-analytic studies. Changing $\beta$ in the range 0.5$\,$-$\,$1 does not significantly alter $P(k)$, hence Eq.~\ref{e:Pbeta_simp} is an excellent approximation for the majority of clusters (Fig.~\ref{fig:beta}, red line). A remarkable feature is that the transition from real to Fourier space does not dramatically deform the profile, in tight analogy with Gaussian functions ($\propto \exp[-k^2]$). 
The spectrum is dominated by the power on large scales, with the core radius playing a crucial role; a progressively rising $r_{\rm c}$ leads to an increase in both the normalization and steepness of the spectrum.

For our study, it is useful to analyze the superposition of the $\beta$-profile and a power-law Kolmogorov noise (with 1D power $\propto k^{-5/3}$), $n_{\rm p}=n_\beta\,(1+\delta)$. Using the convolution theorem, the power spectrum of the perturbed density profile is given by $P(\hat{n}_{\rm p})=P(\hat{n}_\beta)+P(\hat{n}_\beta\ast\hat{\delta})$. The cross terms cancel out since the $\delta$ field is random and the phases are uncorrelated. In Figure \ref{fig:beta}, we show three power spectra: $\beta$-profile (red), noise with $\sim$10 percent relative amplitude (blue), and the superposition of both (black). Beyond the core radius ($k\gta 0.05$), the noise clearly starts to dominate. It is thus not essential to remove the underlying profile or large-scale coherent structures, in order to unveil density perturbations, especially
with substantial turbulence.

\label{lastpage}

\end{document}